\documentclass[unsortedaddress, preprint, nofootinbib, showpacs]{revtex4-1}

\usepackage{amsmath}
\usepackage{amsfonts}
\usepackage{amsthm}
\usepackage{amssymb}
\usepackage{bbm}
\usepackage{graphicx}
\usepackage{hyperref}

	\newcommand{\mrm}{\mathrm}
	\newcommand{\beq}{\begin{equation}}
	\newcommand{\eeq}{\end{equation}}
	\newcommand{\mf}{\mathfrak}
	\newcommand{\td}{\tilde}

	\newcommand{\e}{\mathrm{e}}

\begin{document}

	\title{Deformed phase spaces with group valued momenta}
	\date{\today}
	
	\author{Michele Arzano}
	\email{michele.arzano@roma1.infn.it}
	\author{Francisco Nettel}
	\email{francisco.nettel@roma1.infn.it}
	\affiliation{Dipartimento di Fisica, Universit\`a di Roma La Sapienza,\\ P.le Aldo Moro 5, I-00185 Rome, Italy}

\begin{abstract}
We introduce a general framework for describing deformed phase spaces with group valued momenta. Using techniques from the theory of Poisson-Lie groups and Lie bi-algebras we develop tools for constructing Poisson structures on the deformed phase space starting from the minimal input of the algebraic structure of the generators of the momentum Lie group. The tools developed are used to derive Poisson structures on examples of group momentum space much studied in the literature such as the $n$-dimensional generalization of the $\kappa$-deformed momentum space and the $SL(2, \mathbbm{R})$ momentum space in three space-time dimensions. We discuss classical momentum observables associated to multi-particle systems and argue that these combine according  the usual four-vector addition despite the non-abelian group structure of momentum space.

 \end{abstract}


\maketitle

\section{Introduction}  
\label{sec:intro}

While it is not unusual in physics to contemplate mechanical systems whose configuration space is described by a curved manifold, systems, in particular point particles, with a curved {\it momentum space} have only recently received attention. Interest for momentum space with a non-trivial geometry sparked within the community working on quantum aspects of gravity in the past years and can be traced back to two main lines of research: the study of classical and quantum kinematics of point particles coupled to gravity in three space-time dimensions and Planck-scale {\it deformations} of the Poincar\'e group which can accommodate a fundamental, observer independent, energy scale.

A non-trivial momentum space geometry was first explicitly suggested by 't Hooft in \cite{'tHooft:1996uc}, based on his ``polygon approach" to the kinematics of point-particles coupled to three-dimensional gravity  \cite{thooft1984}. Such particles are described by conical defects of space-time with their mass  proportional to the deficit angle of the cone, the amount a test vector gets rotated when transported around the tip of the cone i.e. the location of the particle. Matschull and Welling \cite{Matschull:1997du} provided the first systematic description of the phase space of such particles showing that their momenta belong to the Lie group $SL(2, \mathbb{R})$, the (double cover of the) three-dimensional Lorentz group. In a parallel work, Bais and Muller \cite{Bais:1998yn} (whose treatment was expanded and carried out in detail in \cite{Bais:2002ye}), starting from the formulation of gravity as a Chern-Simons theory of the Poincar\'e group, argued that a description of the phase space of particles coupled to the theory requires the original symmetry group to be {\it deformed} to a quantum group, known as the ``quantum double" of the Lorentz group. Here, as in \cite{Matschull:1997du}, the key ingredient in the formulation of phase space is the use of  holonomies of the flat connection of the theory, i.e. elements of the Lorentz group, to describe momenta of the particles. In particular these papers provide the first evidence of the connection between quantum deformations of relativistic symmetries and group-valued momenta (here the adjective ``quantum" refers to the fact that ordinary Lie algebra and group structures are replaced by Hopf algebras also known as ``quantum groups").

Quantum deformations of the Poincar\'e algebra were first proposed in \cite{Lukierski:1991pn, Lukierski:1992dt}. The appeal of such models, in particular of the so-called $\kappa$-Poincar\'e algebra, relies on the energy scale set by the deformation parameter $\kappa$, naturally associated with a Planckian scale. These mathematical models served as a basis for the formulation of the so-called ``doubly special relativity" theories \cite{AmelinoCamelia:2000ge,AmelinoCamelia:2000mn}, in which the deformation parameter is regarded as an observer independent scale.  It was soon realized \cite{KowalskiGlikman:2002ft} that the new features of these deformed symmetries can be understood in terms of a non-trivial geometry of momentum space and, in particular, it was showed in \cite{KowalskiGlikman:2004tz} that the $\kappa$-deformed momentum space in four space-time dimensions is given by a Lie group obtained from the Iwasawa decomposition of the de Sitter group $SO(4,1)$. In the last few years the non-trivial geometric properties of momentum space and the associated deformed phase spaces, provided the arena for then new paradigm of  ``relative locality" \cite{AmelinoCamelia:2011bm}.

In all these works the deformed phase spaces are constructed following different techniques tailored for the particular model under consideration. This is especially the case for the fundamental structures defined on such phase spaces: the Poisson brackets. In this paper we describe an approach to the construction of deformed phase spaces with group valued momenta and their Poisson structures which relies only on minimal, model independent, ingredients. Using elements of the theory of Poisson Lie groups we show how it is possible to construct such deformed phase spaces using as the only input the algebraic structure of the generators of the momentum Lie group. 
\footnote{The structures that appear in the theory of Poisson-Lie groups, mainly developed in the seminal papers by Drinfeld \cite{Drinfeld:1983ky, Drinfeld:1986in} Semenov-Tian-Shansky \cite{SemenovTianShansky:1983ik, SemenovTianShansky:1985my} and the classical $r$-matrix introduced previously by Sklyanin \cite{Sklyanin:1980ij, Kulish:1980ii}, came as a classical limit of the structures that appeared in the theory of quantum integrable systems and in turn they appear in the theory of classical integrable systems. See \cite{kosmann1997} for a good review on the subject and the references within.}

Our approach has two advantages: on one side it only relies on the specification of the momemtum group manifold and thus it does not depend on the details of the  fundamental theories from which the model is derived, on the other side it can be easily applied to the construction of deformed phase spaces with {\it any}  momentum Lie group. Moreover it provides a solid starting point for the quantization of models with curved momentum space and their associated deformed symmetries.

The paper is organized as follows: in the next Section we illustrate in detail the basics of the theory of Poisson Lie groups and Lie bi-algebras needed for our discussion. We do so in a self-contained way in order to make the tools borrowed from the mathematical literature accessible to a physics audience. In Section III we re-formulate the conventional phase space description of a relativistic (spinless) point particle using the language of Poisson Lie groups. This will set the stage for the construction, described in Section IV, of the deformed phase space associated with momenta living on the $AN(n)$ Lie group, the $n$-dimensional generalization of the momentum space associated to the $\kappa$-Poincar\'e algebra. In Section V we describe the analogous construction for a phase space with a $SL(2,\mathbbm{R})$ momentum space, the momentum space of point particles coupled to gravity in three space-time dimensions. In the following Section VI we elaborate on the description of phase space for multi-particle systems, a subject which has been object of much controversy in the literature. We conclude in Section VII with a summary of our results and an outlook for future developments.

\section{From symplectic manifolds to Poisson-Lie groups}

We begin our discussion with a review of equivalent formulations of a classical particle's phase space. We start from the conventional picture of phase space as the cotangent bundle of a configuration space equipped with a symplectic form which determines, together with a Hamiltonian and via the Poisson bracket, the dynamics of the system. We show how this familiar picture can be recast in the more abstract language of Poisson-Lie groups and associated $r$-matrices which allow a rather straightforward generalization to phase spaces which {\it do not} possess the structure of cotangent bundles and, in particular, to phase spaces in which momenta belong to a non-abelian Lie group.

\subsection{From symplectic manifolds to Poisson manifolds}
In the usual textbook formulation the states of a classical system belong to a phase space which is given by a {\it symplectic manifold}: an even dimensional differential manifold $\Gamma$ equipped with a non-degenerate closed two-form $\omega$. In most cases $\Gamma$ is the cotangent bundle of the configuration space $M$: $\Gamma = T^*M$. Classical observables are differentiable functions on phase space i.e. belong to $C^\infty (\Gamma)$. The dynamics is determined by a function on $\Gamma$, the Hamiltonian, and the evolution of the system is described by an integral curve of the Hamiltonian vector field $X_H$ on $\Gamma$ determined by Hamilton equations \cite{AbrMars}. One way to describe the action of the Hamiltonian vector field $X_H$ on functions in $C^\infty(\Gamma)$ is in terms of the {\it Poisson bracket}, namely a map $\{\ \cdot\ ,\ \cdot\ \} : C^\infty(\Gamma) \times C^\infty(\Gamma) \to C^\infty(\Gamma)$ with the properties of a Lie bracket, i.e., antisymmetry, the Jacobi identity and the Leibniz-like property. If the Poisson bracket is non-degenerate (there is no point in $\Gamma$ in which $\{f,h\} = 0$ for any $f,h \in C^\infty(\Gamma)$) the Poisson structure is {\it symplectic}. The Hamiltonian vector field $X_f$ can be defined, for any function $f \in C^\infty(\Gamma)$ via the relation
\beq 
\omega(\ \cdot\ ,X_f) = df\,,
\eeq 
from which, the Poisson bracket can be written in terms of the symplectic form as
\beq
\{ f, g \} = - \omega(X_f, X_g) = -X_f(g)\,.
\eeq
The properties which characterize the Poisson bracket are determined by the symplectic form $\omega$, in particular the antisymmetry is given by the two-form nature of $\omega$, the Jacoby identity corresponds to $d\omega = 0$ and the exterior differentiation accounts for the Leibniz-like property. 

Formally we can generalize the mathematical description of the phase space from that of a symplectic manifold to a {\it Poisson manifold}. This is nothing but a pair $(\Gamma, \{\ ,\ \})$ where $\Gamma$ is a differential manifold and  $\{\ ,\ \}$ a Poisson bracket with the properties specified above. The Poisson bracket can be expressed in terms of a skew-symmetric rank-two tensor $w \in T\Gamma \otimes T\Gamma$ by means of the {\it dual pairing} between a vector space and its dual $\langle \ \cdot\ ,\ \cdot\ \rangle : T\Gamma \times T^*\Gamma \to C^\infty(\Gamma)$ and its generalization to tensor fields. Thus, we define the Poisson bracket for $f,g \in C^\infty(\Gamma)$ as
\beq \label{PBbivector}
\{ f,g \} = \langle w, df \otimes dg \rangle.
\eeq
The skew-symmetric tensor $w$ is called {\it Poisson bi-vector} and induces a mapping $w(\; ,\ \cdot\ ): T^*\Gamma \to T\Gamma$ as $\langle w,\ \cdot\ \otimes df \rangle = X_f$ which can also be expressed as $X_f = \{\ \cdot\ , f\}$ \cite{chari}. The important point to notice is that this map is not necessarily invertible and in those cases we do not have a symplectic structure. On the contrary, every symplectic structure is a Poisson structure. 

Our goal is to provide a consistent description of the phase space and Poisson structure of particles whose momenta live on a {\it non-abelian Lie group}. In general for phase spaces which are (direct products of) group manifolds, i.e. where both configuration space and momentum space can be {\it curved spaces}, there is a suitable mathematical formalism to define Poisson brackets which are compatible with the group structure, these are the so-called Poisson-Lie groups. The Poisson structure of a Poisson-Lie group, however, is never symplectic. If one renounces to the requirement of compatibility with the group structure there is still a concise mathematical way to define a symplectic Poisson structure for the Lie group, called the Heisenberg double. Below we provide the minimal concepts needed to consistently determine such symplectic Poisson structure starting from the infinitesimal algebraic structure of the group.

\subsection{Lie groups as phase spaces}
\label{sec:poisson}


Let us start by considering a phase space $\Gamma = T \times G$, given by the Cartesian product of a $n$-dimensional Lie group {\it configuration space} $T$ and a $n$-dimensional Lie group {\it momentum space} $G$. Of course, in this case the phase space no longer bears the structure of a cotangent bundle and it is not obvious how the structures reviewed in the previous section, in particular the Poisson brackets, can be generalized. In order to see how it is possible to extend such tools we start from the Lie algebras associated to these Lie groups, $\mf{t}$ for $T$ and $\mf{g}$ for $G$. Denoting the generators as $\{P_\mu\}$ for $\mf{t}$ and $\{X^\mu\}$ for $\mf{g}$, $\mu = 0,\ldots,n-1$, the Lie brackets are
	\beq  \label{liebrackets}
	[P_\mu, P_\nu] = d_{\mu\nu}^\sigma P_\sigma \qquad \text{and} \qquad [X^\mu, X^\nu] = c^{\mu\nu}_{\sigma} X^\sigma,
	\eeq
where $d_{\mu\nu}^\sigma$ and $c^{\mu\nu}_\sigma$ are the structure constants of the Lie algebras. 

Since $T$ and $G$ will describe, respectively, the positions and momenta of the classical system, it is useful to regard $\mf{t}$ and $\mf{g}$ as {\it dual vector spaces} with a dual pairing defined in terms of the basis elements as
	\beq  \label{dualpairing}
	\langle P_\mu, X^\nu \rangle = \delta_\mu^\nu.
	\eeq
Let us notice that such duality between $\mf{t}$ and $\mf{g}$ allows one to define Poisson brackets on both spaces. To see this let us consider an element  $Y\in \mathfrak{t}$, since $\mf{t}$ is a vector space the tangent space $T_Y\mf{t} \simeq \mf{t}$ is isomorphic to $\mf{t}$ itself.  If we take a smooth function $f\in C^{\infty}(\mf{t})$ then the differential $(df)_Y: T_Y\mf{t}\rightarrow \mathbb{R}$ can be seen as an element of the space $T^*_Y\mf{t} \simeq \mathfrak{g}$.  The Poisson bracket on $C^{\infty}(\mf{t})$ is then given in terms of the commutators of $\mathfrak{g}$ by
\begin{equation}\label{LiePoiss}
\{f,g\}(Y)\equiv \langle Y, [(df)_Y, (dg)_Y]\rangle \,.
\end{equation}
In the same way the Lie brackets on $\mf{t}$ determine Poisson brackets on $C^{\infty}(\mf{g})$. In particular let us consider coordinate functions $f = x^{\mu}$ and $g = x^\nu$ such that $dx^\mu, dx^\nu \in \mf{g}$, it is easy to see that the Lie algebra structure of $\mf{g}$ induces the following Poisson bracket on these functions
\beq
\{x^{\mu}, x^{\nu}\} = c^{\mu\nu}_\sigma x^{\sigma}\,,
\eeq
and, analogously, the Lie algebra structure on $\mf{t}$ defines a Poisson structure \mbox{$\{p_{\mu}, p_{\nu}\} = d_{\mu\nu}^{\sigma} p_{\sigma}$} on $C^{\infty}(\mf{g})$ 
with $p_{\mu}$ coordinate functions on $\mf{g}$ such that the associated differential coincides with the generators $X^{\mu}$. Notice that being such brackets isomorphic to Lie brackets, they automatically satisfy all the required properties for being a Poisson bracket i.e. skew symmetry and Jacobi identity. These brackets are known in the literature as Kirillov-Kostant brackets and are the key object in the description of phase spaces in terms of {\it co-adjoint orbits} \cite{kirillov1976}.	
The Poisson brackets we just discussed reflect a new structure which can be defined on the Lie algebras $\mf{t}$ and $\mf{g}$. Indeed we can write down maps $\delta_{\mf{t}}:\mf{t} \rightarrow \mf{t} \otimes \mf{t}$ and $\delta_{\mf{g}}:\mf{g} \rightarrow \mf{g} \otimes \mf{g}$ given by
	\beq \label{cocommutators}
	\delta_{\mf{t}}(X^\mu) = d_{\alpha\beta}^\mu X^\alpha \otimes X^\beta \qquad \text{and} \qquad \delta_{\mf{g}}(P_\mu) = c^{\alpha \beta}_\mu \ P_\alpha \otimes P_\beta\,. 
	\eeq
It is easy to see that through the dual pairing \eqref{dualpairing} the functions $\delta_{\mf{t}}$ and $\delta_{\mf{g}}$ determine the Lie brackets of $\mf{g}$ and $\mf{t}$, respectively via the relations
\beq
\delta_{\mf{t}} (X^\mu) (P_\alpha, P_\beta) = \langle X^\mu , [P_\alpha , P_\beta]  \rangle\,,\,\,\,\,\,\, \delta_{\mf{g}} (P_\mu) (X^\alpha, X^\beta) = \langle P_\mu , [X^\alpha, X^\beta] \rangle\,,
\eeq
which explains why in the literature such functions are called {\it co-commutators}.

Besides its direct relationship with a Poisson structure on the dual space, the co-commutator defined on a Lie algebra is also linked to a Poisson structure on the associated Lie group. This is of immediate relevance for the analysis we present in this work. In order to illustrate in some detail this link, we review the basics of Poisson structures on a Lie group. The main goal will be to introduce the necessary tools to derive a formula connecting the Poisson bi-vector on the group manifold with the co-commutator on its Lie-algebra.

\subsubsection{Poisson structures on Lie groups}

The framework of Poisson-Lie groups is based on the requirement of having a Poisson structure on the group $G$ such that the group multiplication map $m:G \times G \to G$ denoted as $m(g_1,g_2) = g_1 g_2$ is a {\it Poisson map}. 
This means that, in terms of the Poisson bracket decomposition for functions on the Cartesian product of spaces, the following property must hold:
\begin{multline}
\{f_1 \circ m,f_2 \circ m\}_{G \times G} (g_1,g_2) = \{f_1 \circ m(\ \cdot\ , g_2), f_2 \circ m(\ \cdot\ ,g_2)\}_G (g_1) \\
+ \{f_1 \circ m(g_1,\ \cdot\ ), f_2 \circ m(g_1,\ \cdot\ )\}_G (g_2).
\end{multline}
Such expression can be translated into a condition for the Poisson bracket on the group $\{\cdot, \cdot\}_G$ using the left and right translation maps
\beq
L_{g_1}(g_2) = g_1g_2\,,\,\,\,\, R_{g_1}(g_2)= g_2g_1\,,
\eeq
to become
\beq \label{PoissonProd}
\{f_1, f_2\}_G (g_1g_2) = \{f_1 \circ R_{g_2}, f_2 \circ R_{g_2} \}_G (g_1) + \{f_1 \circ L_{g_1}, f_2 \circ L_{g_1} \}_G (g_2).
\eeq
A Lie group equipped with a Poisson bracket satisfying such property is known as a {\it Poisson-Lie group}.

We would now like to rewrite the Poisson bracket in terms of a Poisson bi-vector. In doing so, we will be able to describe the infinitesimal version of the Poisson structure on the Lie group and to connect it with the co-commutator. 

Let us focus on the right translation map $R_{g}$.  We have the following maps induced by $R_{g_2}$: the pullback between cotangent spaces $R_{g_2}^*: T^*_{g_1g_2}G \to T^*_{g_1}G$ and the pushforward between tangent spaces $R_{g_2}{}_* : T_{g_1}G \to T_{g_1g_2}G$. Indeed, for a tangent vector at $g\in G$, $X \in T_{g_1}G$ and the differential $df \in T_{g_1g_2}G$, we have the dual pairing between elements of the dual spaces
\beq   \label{dualpairingright}
\langle X, R_{g_2}^* df \rangle  = \langle R_{g_2}{}_*X, df \rangle.
\eeq
The analogous relation for the left translation and its induced maps is 
\beq   \label{dualpairingleft}
\langle Y, L_{g_1}^* dh \rangle = \langle L_{g_1}{}_*Y, dh \rangle ,
\eeq
where $Y \in T_{g_1g_2}G$ and $dh \in T^*_{g_2}G$. The pullback and pushforward of $R_{g}$ and $L_{g}$ can be generalized to the tensor product of spaces of any rank using the above relations.

The Poisson bi-vector is then defined in terms of the dual pairing between tangent and cotangent spaces as follows
\beq \label{pbwg}
\{f_1,f_2\} (g) = \langle w_{g}, df_1 \otimes df_2|_{g} \rangle,
\eeq
where for notational simplicity we dropped the subscript $G$ on the Poisson bracket. It is rather straightforward to write down the analogous of condition (\ref{PoissonProd}) for the Poisson bi-vector $w_{g}$ 
\beq \label{bivecg1g2}
w_{g_1g_2} = R_{g_2}{}_*{}^{\otimes 2} |_{g_1} w_{g_1} + L_{g_1}{}_*{}^{\otimes 2} |_{g_2} w_{g_2},
\eeq
which states that a Poisson structure for a Lie group $G$ is Poisson-Lie if and only if the value of its Poisson bivector at $g_1g_2 \in G$ is the sum of the right translate by $g_2$ of its value at $g_1$ plus the left translate by $g_1$ of its value at $g_2$. Notice that for $g_1 = g_2 = e$, where $e \in G$ is the identity element of the group, we have that $w_e = 0$, therefore the rank of the Poisson structure is zero at the identity element of the Lie group, hence {\it the Poisson structure of a Poisson-Lie group is not symplectic}. It is possible, however, to define other Poisson structures for a Lie group which turn the latter into a symplectic manifold and for which the requirement of being Poisson-Lie is dropped. Since these structures result appealing from a physical standpoint, in what follows we show how the co-commutator at the Lie algebra level and the Poisson bi-vector are related. This will allow us to introduce a symplectic Poisson structure on the group and to exhibit the differences between Poisson-Lie and Poisson structures. 

In order to make contact with the notion of co-commutator we focus on the right translate of the Poisson bivector $w$ to the identity element of $G$, denoted by $w^R : G \to \mathfrak{g} \times \mathfrak{g}$, where $\mathfrak{g}= T_e G$ is the Lie algebra of $G$. First, we note that 
\begin{align} \label{wciclo1}
\langle w_g, (df_1 \otimes df_2) |_g \rangle &= \langle w^R(g), R_g^*{}^{\otimes 2} |_e (df_1 \otimes df_2) |_g \rangle, \nonumber \\
&= \langle R_g{}_*{}^{\otimes 2} |_e w^R(g), (df_1 \otimes df_2) |_g \rangle,
\end{align}
hence we have for the Poisson bivector
\beq \label{wciclo2}
w_g = R_g{}_*{}^{\otimes 2} w^R(g)\,.
\eeq

To obtain an expression involving Lie algebra elements we express the group element as $g = e^{tX}$ and consider the derivative of the second term in (\ref{wciclo1}), using $w^R(e) = 0$ we obtain 
\beq \label{diffPB2}
\frac{d}{dt} \langle w^R(e^{tX}), R_{e^{tX}}^* (df_1 \otimes df_2) |_{e^{tX}} \rangle |_{t=0} = \langle \frac{d}{dt} w^R(e^{tX}) \big|_{t=0} , R_{e^{tX}}^*{}^{\otimes 2} (df_1 \otimes df_2) |_{e^{tX}} |_{t=0} \rangle.
\eeq
We now {\it define} the co-commutator in terms of $w^R$ as
\beq  \label{diffPBcoco}
\frac{d}{dt} w^R(e^{tX}) \big|_{t=0} \equiv \delta(X)\,.
\eeq
Denoting $\xi_i = df_i |_{e}$ and equating \eqref{diffPB2} to the derivative of the l.h.s. of \eqref{wciclo1}, we can write down the following relation between co-commutator and Poisson bracket on the group 
\beq  \label{dualrelcoco}
\langle X, d\{f_1,f_2\} |_e \rangle = \langle \delta(X), \xi_1 \otimes \xi_2 \rangle\,.
\eeq

For Poisson-Lie groups $w^R$ must comply with a condition analogous to (\ref{bivecg1g2}) which ensures that the group multiplication on $G$ is a Poisson map. Such condition will translate on a ``local" condition on the co-commutator through (\ref{diffPBcoco}) which we will derive explicitely. Writing
\beq  \label{wciclo3}
w^R(g) = R_{g^{-1}}{}_*{}^{\otimes 2} w_g\,,
\eeq
we can act on \eqref{bivecg1g2} with the tangent linear map associated to the right action $R_{g_1g_2}^{-1} = R_{(g_1g_2)^{-1}} = R_{g_2^{-1}g_1^{-1}}$ and using the identity $R_{g_1}L_{g_2} = L_{g_2}R_{g_1}$ we obtain
\begin{align}  \label{wciclo5}
w^R(g_1g_2) &= R_{g_1^{-1}}{}_*{}^{\otimes 2} L_{g_1}{}_*{}^{\otimes 2} R_{g_2^{-1}}{}_*{}^{\otimes 2} w_{g_2} + R_{g_1^{-1}}{}_*{}^{\otimes 2} w_g  \\
&= R_{g_1^{-1}}{}_*{}^{\otimes 2} L_{g_1}{}_*{}^{\otimes 2} w^R(g_2) + w^R(g_1).
\end{align}
The translation $L_g R_{g^{-1}} g' = g g' g^{-1} = \mathrm{Ad}_g g'$ is the adjoint action of $g$ on $g'$ in $G$. Then, denoting the action of the tangent linear maps as $L_{g_1}{}_* R_{g_1^{-1}}{}_* = \mathrm{Ad}_{g_1}$, we can write the above relation as
\beq  \label{cociclowR}
w^R(g_1g_2) = \mathrm{Ad}_{g_1}{}^{\otimes 2} w^R(g_2) + w^R(g_1),
\eeq
which is the required condition on $w^R$ which ensures that the Poisson structure is compatible with group multiplication. Let us differentiate  \eqref{cociclowR} in order to derive the analogous condition for the co-commutator $\delta(X)$. We start by noticing that from the definition (\ref{diffPBcoco}) it follows\footnote{The property is easily verified by taking the derivative in the direction of $X$ of $w^R(e) = 0$.} that $\delta(-X) = -\delta(X)$.  Next we look at the co-commutator of $[X,Y]$
\begin{align}  \label{dcomm1}
\delta([X,Y]) &= \delta\bigg(\frac{d}{ds}\ \frac{d}{dt} (e^{sX}e^{tY}e^{-sX}) |_{s,t=0}\bigg), \nonumber \\
&= \frac{d}{ds}\ \frac{d}{dt} w^R(e^{sX}e^{tY}e^{-sX}) |_{s,t=0},
\end{align}
taking into account the identity 
\begin{align}  \label{commutator}
[X,Y]= \frac{d}{dt}\ \frac{d}{ds} \bigg(\mathrm{Ad}_{e^{sX}} e^{tY} \bigg) \bigg|_{t,s=0} &= \frac{d}{ds} \bigg(\mathrm{Ad}_{e^{sX}} \bigg) \bigg|_{s=0}\ \frac{d}{dt} e^{tY} \bigg|_{t=0}, \nonumber \\
&= \frac{d}{ds} \bigg( \mathrm{Ad}_{e^{sX}} \bigg) \bigg|_{s=0} Y\,,
\end{align}
and applying the condition (\ref{cociclowR}) for $w^R$ twice we get
\begin{align}  \label{dcomm2}
\delta([X,Y]) &= \frac{d}{ds}\ \frac{d}{dt} \bigg[ w^R(e^{sS}) + \mathrm{Ad}_{e^{sX}}{}^{\otimes 2}\ w^R(e^{tY}e^{-sX}) \bigg]_{s,t=0}, \nonumber \\
&= \frac{d}{ds}\ \frac{d}{dt} \bigg[ w^R(e^{sS}) + \mathrm{Ad}_{e^{sX}}{}^{\otimes 2}\ \big( w^R(e^{tY}) + \mathrm{Ad}_{e^{tY}}{}^{\otimes 2} w^R(e^{-sX}) \big) \bigg]_{s,t=0}, \nonumber \\
&= \frac{d}{ds} \big(\mathrm{Ad}_{e^{sX}}{}^{\otimes 2}\big)\ \frac{d}{dt} w^R(e^{tY}) \big|_{s,t=0} + \frac{d}{ds} \big( \mathrm{Ad}_{e^{sX}}{}^{\otimes 2} \big)\ \frac{d}{dt} \big(\mathrm{Ad}_{e^{tY}}{}^{\otimes 2} \big) w^R(e^{-sX}) \big|_{s,t=0} \nonumber \\
&\hspace{4cm} + \mathrm{Ad}_{e^{sX}}{}^{\otimes 2} \ \frac{d}{dt} \big(\mathrm{Ad}_{e^{tY}}{}^{\otimes 2} \big)\ \frac{d}{ds} w^R(e^{-sX}) \big|_{s,t=0}, 
\end{align}
where we used that $w^R(e) = 0$, $\mathrm{Ad}_{e}{}^{\otimes 2} = \mathbbm{1} \otimes \mathbbm{1}$. Introducing the notation 
\beq \label{XpuntoY}
X.\delta(Y) = (\mathrm{ad}_X \otimes \mathbbm{1} + \mathbbm{1} \otimes \mathrm{ad}_X) \delta(Y)\,,
\eeq
We can rewrite  (\ref{dcomm2}) as the following equation for the co-commutator
\beq \label{cocylecoco}
\delta([X,Y]) = X.\delta(Y) - Y.\delta(X)\,,
\eeq
which is known in the mathematical literature as the {\it co-cycle condition} \cite{chari,tjin1992}.  A Lie algebra equipped with a co-commutator satisfying the co-cycle condition is called a {\it Lie bi-algebra}. We will see that the Lie-bialgebra structure can give rise to Poisson structures which are not Poisson-Lie but are symplectic, and that the requirement \eqref{cocylecoco} for the Lie-bialgebra structure is still crucial in order to have a proper Poisson structure. Thus, from our perspective, the fundamental question is to look for co-commutators which satisfy the co-cycle condition which we can use to construct a Poisson structure on a phase space with group-valued momenta. 

\subsubsection{Poisson structures and the $r$-matrix}
One way to construct a co-commutator which automatically satisfies the co-cycle condition is to consider one of the form
\beq \label{deltar}
\delta(X) \equiv X.r  = \left( \mrm{ad}_X \otimes \mathbbm{1} + \mathbbm{1} \otimes \mrm{ad}_X \right) r\,
\eeq
where $r$ is a generic element of $\mathfrak{g} \otimes \mathfrak{g}$ called the {\it $r$-matrix}. In order for $\delta$ to be a genuine co-commutator the $r$-matrix must satisfy the following two conditions:
\begin{enumerate}
\item The symmetric part of $r$, $r_+ = \frac{1}{2} \left( r^{ij} + r^{ji} \right) X_i \otimes X_j$ is ad-invariant element of $\otimes^2 \mathfrak{g}$, with $r = r^{ij} X_i \otimes X_j$ where $\{X_i\}$ is a basis of $\mathfrak{g}$.
\item $\big[[r,r]\big] = [r_{12}, r_{13}] + [r_{12}, r_{23}] + [r_{13}, r_{23}]$ is an ad-invariant element of $\otimes^3 \mathfrak{g}$,  where
\begin{align}  \label{rij}
r_{12} &= r^{ij}\ X_i \otimes X_j \otimes \mathbbm{1},  \nonumber \\
r_{13} &= r^{ij}\ X_i \otimes \mathbbm{1} \otimes X_j,   \nonumber \\
r_{23} &= r^{ij}\ \mathbbm{1} \otimes X_i \otimes X_j,
\end{align}
and 
\begin{align} 
[r_{12},r_{13}] = r^{ij} r^{kl} [X_i,X_k] \otimes X_j \otimes X_l, \nonumber \\
[r_{12},r_{23}] = r^{ij} r^{kl} X_i \otimes [X_j,X_k] \otimes X_l, \nonumber \\
[r_{13},r_{23}] = r^{ij} r^{kl} X_i \otimes X_k \otimes [X_j,X_l]. 
\end{align}
\end{enumerate}
The first condition is directly related to skew-symmetry of the Poisson bracket and Lie brackets defined by $\delta$. Analogously, the second condition ensures that such brackets satisfy the Jacobi identity. Notice that from the first condition we see that $(\mathrm{ad}_g \otimes \mathbbm{1} + \mathbbm{1} \otimes \mathrm{ad}_g) r_+ = 0$ is trivially satisfied if $r_+=0$, i.e. using a skew-symmetric $r$-matrix. It is also clear that the simplest way of satisfy the second condition is if $\big[[r,r]\big] = 0$. This equation is know as the {\it classical Yang Baxter equation} (CYBE) and its solution is called the {\it classical $r$-matrix}. The condition of $\big[[r,r]\big]$ being ad-invariant
\beq  \label{adschouten}
X^i . \big[[r,r]\big] = 0 \quad \forall X^i \in \mathfrak{g}
\eeq
where
\beq  \label{adschoutenop}
X^i . \big[[r,r]\big] \equiv \big(\mathrm{ad}_{X^i} \otimes \mathbbm{1} \otimes \mathbbm{1}\ +\  \mathbbm{1} \otimes \mathrm{ad}_{X^i} \otimes \mathbbm{1}\ +\ \mathbbm{1} \otimes \mathbbm{1} \otimes \mathrm{ad}_{X^i}  \big) \big[[r,r]\big],
\eeq
is known as the {\it modified classical Yang-Baxter equation} (mCYBE). The most important point for us is that the co-commutator, as defined by   \eqref{deltar}, can be \textit{integrated} to a Poisson structure on $G$. One possibility is to hold on the requirement of compatibility with group multiplication and obtaining a Poisson-Lie structure on $G$. Nevertheless, other Poisson structures can be associated to the same co-commutator which are not Poisson-Lie but which e.g. are symplectic and thus are good candidates for describing the Poisson bracket of a deformed phase space. These last structures are the ones we are interested in.

Before moving on let us describe in some detail the Poisson bi-vectors associated a $r$-matrix. A possible choice of Poisson bi-vector associated to a given $r$-matrix is such that its right translate to the identity element of $G$ is given by 
\beq \label{wRr} 
w^R(g) = \mathrm{Ad}_g{}^{\otimes 2} r - r.
\eeq
Writing an element of $G$ as $g = e^{tX}$, is a straightforward calculation to check that \eqref{wRr} has the ``correct" derivative
\begin{align} \label{derivwR}
\delta(X) &= \frac{d}{dt} w^R(e^{tX}) \Big|_{t=0}, \nonumber \\
&= \bigg[ \frac{d}{dt} \big(\mathrm{Ad}_{e^{tX}} \big) \otimes \mathrm{Ad}_{e^{tX}} + \mathrm{Ad}_{e^{tX}} \otimes \frac{d}{dt} \big(\mathrm{Ad}_{e^{tX}} \big) \bigg]_{t=0} \ r, \nonumber \\
& = ( \mathrm{ad}_X \otimes \mathbbm{1} + \mathbbm{1} \otimes \mathrm{ad}_X )\ r = X.r\,.
\end{align}
It is also easily checked that (\ref{wRr}) satisfies the co-cycle property \eqref{cociclowR}.

It is possible to define a different Poisson structure on $G$, starting from the same $r$-matrix, which also satisfies the above properties. This structure is determined by the Poisson bi-vector
\beq \label{wRrHeisenberg}
w^R(g) = \mathrm{Ad}_g{}^{\otimes 2} r + r^*,
\eeq
where $r^*$ is minus the transpose of $r$, that is, if $r = r^{ij} X_i \otimes X_j$ then $r^* = -r^{ji} X_i \otimes X_j$,  ($r^* = - r^t$). As we will see this Poisson bivector {\it does not} give rise to a Poisson-Lie structure but it allows to define a symplectic Poisson structure on the group manifold as it is not necessarily degenerated at the group identity.

It is useful to write down the explicit form of the Poisson brackets on the group in terms of the $r$-matrix.  The Poisson bracket associated to \eqref{wRr} can be obtained from  \eqref{pbwg} 
written in terms of the right translate of $w_g$ to the identity element, c.f. equation \eqref{wciclo1}
\begin{align}  \label{wgPBr}
\{f_1,f_2\}(g) &= \langle w^R(g), R_g^*{}^{\otimes 2}\ (df_1 \otimes df_2)|_g \rangle, \nonumber \\
&= \langle (\mathrm{Ad}_g^{\otimes 2}\ r - r, R_g^*{}^{\otimes 2}\ (df_1 \otimes df_2)|_g \rangle, \nonumber \\
&= \langle L_g{}_* R_{g^{-1}}{}_*{}^{\otimes 2} \ r, R_g^*{}^{\otimes 2}\ (df_1 \otimes df_2)|_g \rangle \nonumber \\
&\hspace{4cm} - \langle  r, R_g^*{}^{\otimes 2}\ (df_1 \otimes df_2)|_g \rangle, \nonumber \\
&= \langle r, L_g^*{}^{\otimes 2} \ (df_1 \otimes df_2)_g \rangle  - \langle r, R_g^*{}^{\otimes 2}\ (df_1 \otimes df_2)|_g \rangle\,.
\end{align}
where in the third line we used the equality $L_{g_1}{}_* R_{g_1^{-1}}{}_* = \mathrm{Ad}_{g_1}$. Thus we have 
\beq
\{f_1,f_2\}(g) =  \langle r, L_g^*{}^{\otimes 2} \ (df_1 \otimes df_2)_g \rangle \nonumber - \langle r, R_g^*{}^{\otimes 2}\ (df_1 \otimes df_2)|_g \rangle\,.
\eeq
The corresponding expression for the bivector  \eqref{wRrHeisenberg} is given by
\beq  \label{wgPBHeisenberg}
\{f_1, f_2\}(g) = \langle r, L_g^*{}^{\otimes 2} (df_1 \otimes df_2) |_g \rangle + \langle r^*, R_g^*{}^{\otimes 2} (df_1 \otimes df_2 ) |_g \rangle.
\eeq
In order to write the brackets in a form which will be more convenient for actual calculations we express the $r$-matrix as $r = r^{ij}\ X_i \otimes X_j$ in terms of a basis $\{X_i\}$ for $\mathfrak{g} = T_e G$. The bracket  \eqref{wgPBr} will be written as 
\begin{align} \label{PBrij}
\{f_1,f_2\} (g) &= \langle r^{ij}\ X_i \otimes X_j, \big( L_g^*{}^{\otimes 2}  - R_g^*{}^{\otimes 2} \big) |_e\ (df_1 \otimes df_2)_g \rangle, \nonumber \\
&= \langle r^{ij}\ \big( L_g{}_*{}^{\otimes 2}\ (X_i \otimes X_j) - R_g{}_*{}^{\otimes 2}\ (X_i \otimes X_j), (df_1 \otimes df_2)_g \rangle, \nonumber \\
&= \langle r^{ij}\ \big( X^L_i \otimes X^L_j - X^R_i \otimes X^R_j \big), (df_1 \otimes df_2)_g \rangle
\end{align}
where $X^L_i = L_g{}_*\ X_i$ and $X^R_i = R_g{}_*\ X_i$ are the left and right translate of $X_i \in T_e G = \mathfrak{g}$, and thus in compact form
\beq  \label{PBleftright}
\{f_1,f_2\} = r^{ij}\ \big( X^L_i f_1\ X^L_j f_2 - X^R_i f_1\ X^R_j f_2 \big)\,.
\eeq
Analogously for the bracket  \eqref{wRrHeisenberg} we can write 
\begin{align}  \label{PBleftrightHei}
\{f_1,f_2\} &= r^{ij}\ X^L_i f_1\ X^L_j f_2 + (r^*)^{ij} X^R_i f_1\ X^R_j f_2 \nonumber \\
&= r^{ij}\ X^L_i f_1\ X^L_j f_2 - r^{ji} X^R_i f_1\ X^R_j f_2.
\end{align}
Let us finally notice that if $G$ is a matrix group, the matrix elements $t_{lm}$ in $GL_n(\mathbb{R})$ can be seen as coordinate functions on the group, $t_{lm}(g)$. The Poisson bracket is determined by these matrix elements
\beq
X^L(t_{lm}) = (TX)_{lm}, \qquad X^R(t_{lm}) = (XT)_{lm},
\eeq
where $T$ is the matrix whose elements are $t_{lm}$. Then
\beq
\{t_{mn}, t_{kl}\} = \sum_{a,b} \left( r_{an\, bl}\ t_{ma} t_{kb} - r_{makb} t_{an} t_{bl}\right),
\eeq
where $r_{an\, bl} \equiv (r^{ij} X_i \otimes X_j)_{an\,bl} = r^{ij}\ (X_i)_{an} (X_j)_{bl}$.

Usually, this Poisson bracket is denoted as $\{t_{mn}, t_{kl} \} = \{ T\ \overset{\otimes}{,}\ T \}_{mnkl}$ and it is written as
\beq
\{T\ \overset{\otimes}{,}\ T\} = [ T \otimes T, r ].
\eeq
In the next section we will apply the tools illustrated so far to cartesian product Lie groups of the type $T \times G$. Imposing a duality relation between the Lie algebras $\mf{t}$ and $\mf{g}$ the group manifold $\Gamma = T \times G$ can be seen as a deformation of ordinary phase spaces with group manifold configuration space {\it and} momentum space.

\subsection{Deforming phase spaces: the classical doubles}

In order to define a Poisson structure on the group phase space $\Gamma = T \times G$ we will look for an ``exponentiated" version of co-commutators defined on the Lie (bi)-algebra $\mf{t} \oplus \mf{g}$. The starting point, of course, will be to define a Lie algebra structure on the vector space $\mf{t} \oplus \mf{g}$ compatible with the duality relation (\ref{dualpairing}) between $\mf{t}$ and $\mf{g}$.  Such relation is encoded in the natural inner product on $\mf{t} \oplus \mf{g}$ given by 
      \beq \label{extdualpairing}
	(P_\mu, P_\nu) = 0, \qquad (X^\mu, X^\nu) = 0 \qquad \text{and} \qquad (P_\mu, X^\nu) = \langle P_\mu, X^\nu \rangle\,.
	\eeq
We want to define on $\mf{t} \oplus \mf{g}$ a Lie bracket such that the inner product above is {\it invariant} under the adjoint action of the elements of $\mf{t} \oplus \mf{g}$, that is
\beq  \label{adinvariantinner}
	([Z_A,Z_B],Z_C) = (Z_A,[Z_B,Z_C]),
\eeq
where $Z_A = \{P_\mu, X^\mu\}$, $A = 1,\ldots 2n$. The following Lie brackets 
\beq  \label{liealgdouble}
	[P_\mu, P_\nu] = d_{\mu\nu}^\sigma P_\sigma, \qquad [X^\mu, X^\nu] = c^{\mu\nu}_\sigma X^\sigma \qquad \text{and} \qquad [P_\mu,X^\nu] = c^{\nu\sigma}_\mu P_\sigma - d_{\mu\sigma}^\nu X^\sigma,
	\eeq
comply with such requirement as it can be easily verified. However, in order to show that these brackets turn $\mf{t} \oplus \mf{g}$ into a Lie algebra we must ensure that they satisfy the Jacobi identity. It turns out that the brackets (\ref{liealgdouble}) on $\mf{t} \oplus \mf{g}$ satisfy the Jacobi identity {\it if and only if} the co-commutator $\delta_{\mf{t}}$ on the Lie algebra $\mf{t}$ satisfies the co-cycle condition (\ref{cocylecoco})  \cite{chari}. As we discussed at the beginning of Section II.B, the Lie algebra structure on $\mf{t}$ defines a co-commutator $\delta_{\mf{g}}$ on the dual ``momentum" Lie algebra $\mf{t}$ via 
\beq \label{LBdualcoco}
\langle P, [X_i, X_j]_{\mathfrak{t}} \rangle = \langle \delta_{\mathfrak{g}}(P), X_i \otimes X_j \rangle\,.
\eeq
Thus we see that given the Lie algebra structure \eqref{liealgdouble} on $\mf{t} \oplus \mf{g}$ and the pairing through the inner product \eqref{extdualpairing}, a Lie bi-algebra structure on $\mf{t}$ (and by duality on $\mf{g}$) is naturally induced and it can be shown to be {\it unique}  \cite{chari}.

We want to define now a Lie-bialgebra structure on the whole direct sum Lie algebra $\mathcal{D} = \mf{t} \oplus \mf{g}$ i.e. define a co-commutator $\delta_{\mathcal{D}}$ which reproduces $\delta_{\mf{t}}$ and $\delta_{\mf{g}}$ as given in equations \eqref{cocommutators} when restricted, respectively, to $\mf{t}$ and $\mf{g}$. It turns out that there is a canonical way\footnote{Sometimes the $r$-matrix is written directly as the skew-symmetric part $r_- = \frac{1}{2}\left( P_\mu \otimes X^\mu - X^\mu \otimes P_\mu \right) \equiv P_\mu \wedge X^\mu$, which obviously yields the same Lie-bialgebra structure. One just have to notice that in such cases the $r_-$-matrix satisfies trivially $\mrm{ad} r_+ = 0$ and also it is a solution for the mCYBE instead of the CYBE. See \cite{Zakrzewski1994} for the use of an analogous classical $r$-matrix in the context of deformations of the Poincar\'e group.} of defining such co-commutator in terms of the $r$-matrix belonging to $\mf{t} \otimes \mf{g}$
\beq \label{rmadouble}
r = P_\mu \otimes X^{\mu}\,, 
\eeq
as 
\beq 
\label{cocodouble}
	\delta_{\mathcal{D}}(Z_A) = Z_A.r = \left( \mrm{ad}_{Z_A} \otimes \mathbbm{1} + \mathbbm{1} \otimes \mrm{ad}_{Z_A} \right) r\,.
\eeq
It is easily verified that such co-commutator reduces to  \eqref{cocommutators}, when $Z_A= X^{\mu}$ or $Z_A= P_{\mu}$. It can be also proved \cite{chari} that \eqref{cocodouble} defines a genuine Lie bi-algebra structure on $\mathcal{D}$, i.e. that $\delta_{\mathcal{D}}(Z_A)$ complies with the properties listed in Section IIB.

We can now use the $r$-matrix \eqref{rmadouble} do define a Poisson structure along the lines illustrated in the previous section. There we saw that, given an appropriate $r$-matrix, there are two possible choices of Poisson bi-vector which can be used to define a Poisson structure on $\Gamma = T \times G$. Let us consider two functions on $\Gamma$, $f_1,f_2 \in C^\infty(T \times G)$, then the Poisson bracket is given by $\{ f_1, f_2 \}(h) = \langle w_h, (df_1 \otimes df_2)_h \rangle$, for $h \in T \times G$. In terms of the $r$-matrix we have a Poisson bracket given by 
	\beq \label{rmatrix1}
	\{f_1, f_2\} = - r^{AB} \big( Z_A^R f_1\, Z_B^R f_2\ \pm\ Z_A^L f_1\, Z_B^L f_2   \big),
	\eeq
where $Z_A^L = L_h{}_* Z_A$ and $Z_B^R = R_h{}_* Z_B$ are the left and right translates of $Z_A \in \mf{t} \oplus \mf{g}$. If $\Gamma$ is a matrix group then \eqref{rmatrix1} is given as
	\beq \label{rmatrix2}
	\{\gamma_{ij}, \gamma_{kl}\} = - \sum_{a,b} ( r_{iakb} \gamma_{aj} \gamma_{bl} \pm r_{ajbl} \gamma_{ia} \gamma_{kb} ),
	\eeq
where $\gamma \in \Gamma$ with $\gamma_{ij}$ its components which can be understood as coordinate functions for the group, \mbox{$r_{iakb} \equiv (r^{AB} Z_A \otimes Z_B)_{iakb}$}, and $Z_A^L(\gamma_{ij}) = (\gamma Z_A)_{ij}$, $Z_A^R(\gamma_{ij}) = (Z_A \gamma)_{ij}$. This can be expressed compactly as
	\beq \label{rmatrix3}
	\{ \gamma\ \overset{\otimes}{,}\ \gamma \} = - [r, \gamma \otimes \gamma]_\pm,
	\eeq
with the plus subscript denoting the anti-commutator and the minus for the commutator. The bracket given by the commutator equips $\mathcal{D}$ with the structure of {\it Drinfeld double} and the one with the anticommutator the strucuture of the {\it Heisenberg double}. We would like to stress the fundamental difference between these two structures. On the one hand, the Drinfeld structure \eqref{wRr} satisfies the compatibility condition \eqref{bivecg1g2} for having a Poisson-Lie group, but the Poisson structure is always degenerated at the identity element $e \in G$, hence the structure is not symplectic\footnote{It is possible, however, to foliate the group manifold in a set of {\it symplectic leaves} with the Poisson structure of the manifold restricted to each leaf \cite{alekseev1994}.}. On the other hand, in the case of the Heisenberg double, one renounces to the requirement of compatibility between the group multiplication and the Poisson structure favouring the possibility of a global symplectic structure \cite{kosmann1997}. 
In what follows we will focus on explicit examples of phase spaces and their Poisson structures making use of the tools developed so far. We will start  from a ordinary massive relativistic particle and will proceed to consider {\it deformed phase spaces} in three and more space-time dimensions which are characterized by momenta living on a non-abelian Lie group.

\section{Relativistic spinless particle: flat momentum space}
\label{subsec:flat}

In this section we start by describing the (undeformed) phase space of a relativistic spinless particle using the tools developed in the previous Section.
Even though the following treatment might appear as a byzantine academic exercise, it will serve as a starting point for introducing the deformations of momentum space which we will develop in the next Sections.

The configuration space of a spinless relativistic particle in $n+1$-dimensions can be identified with the Abelian group of translations, $\mathcal{T} \simeq \mathbb{R}^{n,1}$. At any point in the configuration space $x \in \mathcal{T}$ the cotangent (momentum) space is $T_x^* \mathbb{R}^{n,1} = \mathbb{R}^{n,1}{}^*$, where $\mathbb{R}^{n,1}{}^*$ stands for the dual, as a vector space, to $\mathbb{R}^{n,1}$. Following a ``geometric" approach, at this point one would introduce a symplectic structure and with it a Poisson bracket for the space of functions on the cotangent bundle $T^*\mathcal{T}$. Here we will follow an algebraic approach along the lines discussed in the previous section. Our phase space manifold $\Gamma$ is given by
	\beq
	\Gamma = \mathcal{T} \times \mathcal{T}^* \simeq \mathbb{R}^{n,1} \times \mathbb{R}^{n,1}{}^*.
	\eeq

We denote the Lie algebra associated to each component of $\Gamma$ as $\mf{t}$ and $\mf{t}^*$ for $\mathcal{T}$ and $\mathcal{T}^*$, respectively, and the coordinates for $\mathcal{T}$ as $x^\mu$ and $p_\mu$ for $\mathcal{T}^*$, with $\mu=0,\ldots, n$. For the Lie algebra $\mathfrak{t}$ we denote the basis elements as $\{P_\mu \}$, whereas for $\mathfrak{t}^*$ we use $\{X^\mu \}$. The (trivial) Lie brackets are
	\beq  \label{trivialcomm}
	[P_\mu,P_\nu] = 0 \qquad \text{and} \qquad [X^\mu, X^\nu] = 0,
	\eeq
for all $\mu,\nu$. We will see that the coordinate bases for the Lie algebras related to the groups coordinates are 
	\beq  \label{coordbases}
	P_\mu = \frac{\partial}{\partial x^\mu} \qquad \text{and} \qquad X^\mu = -\frac{\partial}{\partial p_\mu},
	\eeq
where $X^\mu = \eta^{\mu\nu} X_\nu$, $\eta^{\mu\nu} = \mrm{diag}(+,-,\ldots,-)$ and $a,b = 1,\ldots,n$.
Taking into account that $\mathfrak{t}$ and $\mathfrak{t}^*$ are dual spaces $\langle P_\mu, X^\nu \rangle = \delta_\mu^{\; \nu}$, we can define an inner product for $\mathfrak{t} \oplus \mathfrak{t}^*$ extending the dual pairing as follows
	\beq  \label{innerprodD}
	( P_\mu, P_\nu ) = ( X^\mu, X^\nu ) = 0 \qquad \text{and} \qquad ( P_\mu, X^\nu ) = \langle P_\mu, X^\nu \rangle = \delta_\mu^{\; \nu}.
	\eeq
As we saw in the previous section we can define a Lie bracket for $\mathfrak{t} \oplus \mathfrak{t}^*$ asking that the inner product of the direct sum is $\mathrm{ad}$-invariant. The ``mixed'' commutator vanishes for any two members of $\mathfrak{t} \oplus \mathfrak{t}^*$, thus we obtain an Abelian Lie algebra
	\beq  \label{commD}
	[P_\mu, P_\nu] = 0, \qquad [X^\mu, X^\nu] = 0, \qquad \text{and} \qquad [P_\mu, X^\nu] = 0.
	\eeq
Therefore, for $\mathfrak{t}$ and $\mathfrak{t}^*$ we have trivial co-commutators
	\beq \label{coco-t}
	\delta_{\mathfrak{t}}(P_\mu) = 0 \qquad \text{and} \qquad \delta_{\mathfrak{t}^*}(X^\mu) = 0,
	\eeq
in accordance with the two first expressions in \eqref{commD}.

At this point we only have a Lie algebra structure for $\mathfrak{t} \oplus \mathfrak{t}^*$. We are interested in defining a Lie-bialgebra structure on such Lie algebra in order to obtain the ``double" of the Lie algebra of translations, $\mathcal{D}(\mathfrak{t})$. Since we are dealing with an abelian Lie algebra, the Lie bialgebra structure will have a trivial co-commutator. Nevertheless, we want to carry on with this construction as we want to present the next cases as a \emph{deformation} of momentum space to a non-abelian Lie group.

The co-commutator for the double $\mathcal{D}(\mathfrak{t})$ can be obtained from \eqref{cocodouble} choosing an $r$-matrix which fulfills the conditions of the ad-invariance of its symmetric part and the (modified) Yang Baxter equation. In this case where the Lie algebra of $\mathfrak{t} \oplus \mathfrak{t}^*$ is trivial any $r$-matrix satisfies the Lie-bialgebra conditions and has a trivial co-commutator. The canonical Poisson bracket can be obtained from the following anti-symmetric $r$-matrix
	\beq  \label{rcanonical}
	r = \frac{1}{2} \left( P_\mu \otimes X^\mu - X^\mu \otimes P_\mu \right) \equiv \, P_\mu \wedge X^\mu .
	\eeq
Grouping the set of generators for $\mathfrak{t}$ and $\mathfrak{t}^*$ as $Z^A = (P_\mu, X^\mu)$ where $A=1,\ldots, 2(n+1)$ we see that 
	\beq \label{cocoDtrivial}
	\delta_{\mathcal{D}(\mathfrak{t})} (Z^A) = (\mathrm{ad}_{Z^A} \otimes \mathbbm{1} + \mathbbm{1} \otimes \mathrm{ad}_{Z^A}) r = 0
	\eeq
for all $Z^A \in \mathfrak{t} \oplus \mathfrak{t}^*$, i.e. we have Lie-bialgebra structure turning $\mathfrak{t} \oplus \mathfrak{t}^*$ in a classical double. 

We can now derive the Poisson bracket for the Poisson-Lie group whose Lie algebra is given by $\mathfrak{t} \oplus \mathfrak{t}^*$ with the trivial commutators \eqref{commD}. From the canonical $r$-matrix \eqref{rcanonical} and choosing the plus sign in \eqref{rmatrix1} we see that
	\begin{align}
	\{f_1, f_2\} &=- 2 r^{AB} Z_A f_1 Z_B f_2, \nonumber \\
	&= -\delta^\mu_\nu \left(  P_\mu f_1 X^\nu f_2 -  X^\nu f_1 P_\mu f_2 \right).
	\end{align}
Using the coordinate basis \eqref{coordbases} we obtain 
	\beq
	\{f_1, f_2\} = \frac{\partial f_1}{\partial x^\mu} \frac{\partial f_2}{\partial p_\mu} - \frac{\partial f_1}{\partial p_\mu} \frac{\partial f_2}{\partial x^\mu} ,
	\eeq
and the Poisson brackets for the coordinate functions on the phase space are 
	\begin{align}  \label{pbflat}
	\{x^0, x^a\} = 0 \qquad  \{x^a, x^b\} &= 0 \qquad \{p_0,p_a\} = 0 \qquad \{p_a, p_b\} = 0 \nonumber \\
	\{x^0, p_0\} = 1 \qquad \{x^a, p_0 \} &= 0 \qquad \{x^0, p_a\} = 0 \qquad \{x^a, p_b\} = \delta^a_b,
	\end{align}
where $a,b = 1,\ldots,n$. This Poisson structure is also symplectic as the Poisson bivector is non-degenerate. Notice that if we had chosen the minus sign in \eqref{rmatrix1}, i.e. the Drinfeld double structure, we would have obtained a trivial Poisson-Lie structure! Thus the Heisenberg double structure, in the flat momentum space, can be used to define the canonical Poisson brackets of textbook classical mechanics of a relativistic point particle.

\section{Deforming momentum space to the $AN(n)$ group}
\label{sec:deform}

As a first example of non-abelian momentum space we will consider an $n$-dimensional Lie sub-group of the $n+2$-dimensional Lorentz group $SO(n+1,1)$ denoted as $AN(n)$. The three and four-dimensional versions of this group have been subject of an extensive study over the past  years in the context of $\kappa$-deformations of relativistic symmetries \cite{KowalskiGlikman:2003we, Freidel:2007hk, meusburger2009, Arzano:2010kz, arzano2011, AmelinoCamelia:2011nt}. These models are characterized by {\it deformed} translation generators which act according to a ``generalized Leibniz rule" on tensor product representations. Such generators are associated to a $AN(n)$ momentum space which as a manifold is described by ``half" of the $n+1$-dimensional de Sitter space. The deformation parameter $\kappa$, with dimension of inverse length, is related to the curvature of de Sitter space. The Lie algebra $\mf{an}(n)$, when its generators are identified with space-time coordinates, is known in the literature as the $n+1$-dimensional $\kappa$-Minkowski space. For further technical details on $\kappa$-deformations the interested reader can consult \cite{Majid:1994cy, Kosinski:1999ix, AmelinoCamelia:2001fd, Agostini:2006zza, Arzano:2007ef, Arzano:2009ci, Kim:2009jk, Meljanac:2010ps, Borowiec:2010yw, Arzano:2014jfa}.  The main goal of this Section will be to show that starting from {\it minimal} ingredients, namely the structures constants the Lie algebra $\mf{an}(n)$, we can construct a suitable Poisson structure on a deformed phase space in which momenta belong to the $AN(n)$ Lie group. Such phase space provides a description of the kinematics of a classical $\kappa$-deformed relativistic particle.

Our starting point will be the manifold $\Gamma = T \times AN(n)$ where $T= \mathbbm{R}^{n,1}$ that is the ordinary $n+1$-dimensional Minkowski configuration space. The usual flat momentum space, however, is now replaced by the group manifold $AN(n)$. In all examples that will follow we will restrict to models with flat configuration space $T= \mathbbm{R}^{n,1}$.

Let us look at the Lie algebras of both components of the cartesian product group $\Gamma$. Denoting again with $\{P_\mu\}$ the basis of $\mf{t}$ and with $\{\td{X}^\mu\}$, $\mu = 0, \ldots, n$ the basis of $\mf{an}(n)$ we have 
	\beq \label{liealgdeform}
	[P_\mu, P_\nu] = 0 \qquad \text{and}  \qquad [\td{X}^\mu, \td{X}^\nu] = - \frac{1}{\kappa} \left( \td{X}^\mu \delta^\nu_0 - \td{X}^\nu \delta^\mu_0 \right),
	\eeq
We immediately see that in the limit $\kappa \to \infty$ we recover the undeformed case of a ordinary relativistic particle reviewed in the previous Section. The algebra $\mf{an}(n)$ is usually expressed as
	\beq \label{liealgan}
	[\tilde{X}^0, \tilde{X}^a] = \frac{1}{\kappa} \tilde{X}^a \qquad [\tilde{X}^a, \tilde{X}^b] = 0,
	\eeq
where $a,b = 1, \ldots, n$. The two vector spaces $\mf{t}$ and $\mf{an}(n)$ are dual with respect to the inner product $\langle P_\mu, \td{X}^\nu \rangle = \delta_\mu^\nu$. The Lie algebra structure of $\mf{an}(n)$ reflects on a non-trivial co-commutator for $\mf{t}$
	\beq \label{cocottilde}
	\delta_{\mathfrak{t}} (P_\mu) = -\frac{1}{\kappa} \left( P_\mu \otimes P_0 - P_0 \otimes P_\mu \right) = \frac{2}{\kappa} \left( P_0 \wedge P_\mu \right),
	\eeq
while for $\mf{an}(n)$ we have
	\beq \label{cocottildestar}
	\delta_{\mathfrak{an}} (\tilde{X}^\mu) = 0.
	\eeq

The direct sum of Lie algebras $\mf{t} \oplus \mf{an}(n)$ can be equipped with an inner product, cfr. \eqref{adinvariantinner}, invariant under the action of $\mf{t}$ and $\mf{an}(n)$
	\beq \label{innerprod}
	(P_\mu, P_\nu) = 0, \qquad (\td{X}^\mu, \td{X}^\nu) = 0 \qquad \text{and} \qquad (P_\mu, \td{X}^\nu) = \delta_\mu^\nu\,.
	\eeq
Such product can be used to derive Lie brackets defining a Lie algebra structure on $\mf{t} \oplus \mf{an}(n)$. The Lie brackets are given by \eqref{liealgdeform} together with
	\beq \label{lietan}
	[P_\mu, \tilde{X}^\nu] = -\frac{1}{\kappa} \left(\delta^\nu_\mu P_0 - \delta^\nu_0 P_\mu \right),
	\eeq
which written explicitly read
	\begin{align}
	[P_0, \tilde{X}^{\mu}] &= 0 \qquad [P_a, \tilde{X}^0] = \frac{1}{\kappa}  P_a \qquad [P_a, \tilde{X}^b] = -\frac{1}{\kappa}  \delta_a^b P_0.
	\end{align}

As we showed in Section II, in order to define a Poisson structure on $\Gamma$ is suffices to introduce an $r$-matrix which turns $\mf{t} \oplus \mf{an}(n)$ into a Lie bi-algebra. A candidate $r$-matrix can be obtained from \eqref{rcanonical} simply replacing $ X^\mu$ with the new generators  $\tilde{X}^\mu$
	\beq  \label{rmatrixan}
	r = \frac{1}{2} \left(P_\mu \otimes \tilde{X}^\mu - \tilde{X}^\mu \otimes P_\mu \right) \equiv P_\mu \wedge \tilde{X}^\mu.
	\eeq
It is easily checked that this skew-symmetric $r$-matrix \eqref{rmatrixan} satisfies the two conditions to render $\mf{t} \oplus \mf{an}(n)$ a Lie-bialgebra, that is, $r_+ = 0$ is trivially ad-invariant and a direct calculation shows that the Schouten bracket satisfies the modified Classical Yang-Baxter equation 
	\beq  \label{mcybetilde}
	Z_A. \big[ [r,r]\big] = 0 \qquad \forall \ Z_A \in \mathfrak{t} \oplus \mathfrak{an}(n).
	\eeq
The co-commutator on $\mf{t} \oplus \mf{an}(n)$ is defined by \eqref{cocodouble} and on the generators of $\mathfrak{t}$ and $\mathfrak{an}(n)$ reduces to
	\begin{align}  \label{coco-Dtilde}
	\delta_{\mathfrak{t} \oplus \mathfrak{an}} (P_\mu)  &= - \frac{1}{\kappa} \left( P_\mu \otimes P_0 - P_0 \otimes P_\mu \right) = \frac{2}{\kappa}\, P_0 \wedge P_\mu, \\
	\delta_{\mathfrak{t} \oplus \mathfrak{an}} (\tilde{X}^\mu) &= 0\,.
	\end{align}
It can be verified that the co-commutator satisfies the cocycle condition
	\beq  \label{cocycle-Dtilde}
	\delta_{\tilde{\mathcal{D}}(\mathfrak{t})} \big([Z_A,Z_B]\big) = Z_A .\ \delta_{\tilde{\mathcal{D}}(\mathfrak{t})}(Z_B) - Z_B .\  		\delta_{\tilde{\mathcal{D}}(\mathfrak{t})}(Z_A)\,,
	\eeq
so the Lie-bialgebra $\mathfrak{t} \oplus \mathfrak{an}(2)$ can be seen as a {\it classical double} $\tilde{\mathcal{D}}(\mathfrak{t}) \equiv \mathfrak{t} \oplus \mf{an}(n)$.
	
We can use the structures just described to construct a Poisson structure on the group $\td{D}(T) = T \times AN(n)$. We will use a matrix representation for the Lie algebra $\mf{an}(n)$ and we will extend it in order to include in the representation the Lie algebra $\mf{t}$. It is common practice to describe the group $AN(n)$ in embedding coordinates which make clear the identification of the group manifold with half of the $(n+1)$-de Sitter hyperboloid embedded in $(n+1) +1$-Minkowski space. In this case the generators of the corresponding Lie algebra are given by combinations of the generators of the Lie algebra $\mf{so}(n+1,1)$ and are represented by $(n+2) \times (n+2)$ matrices \cite{Arzano:2014jfa}. However, in order to include translations, to obtain a matrix representation for the Lie algebra $\mf{t} \oplus \mf{an}(n)$, it will be necessary to work in a different representation. We thus introduce the {\it adjoint representation} $\mathcal{R}$ for $\mf{an}(n)$ defined by $\mathcal{R}: \mf{an}(n) \to \mf{gl}(\mf{an}(n))$, $\td{X} \mapsto \mrm{ad}_{\td{X}}$, where $\mrm{ad}_{\td{X}}(\td{Y}) := [\td{X},\td{Y}]$ for $\td{X},\td{Y} \in \mf{an}(n)$. The generators of the adjoint representation are determined by structure constants of the Lie algebra, $[\td{X}^\mu, \td{X}^\nu] = c^{\mu\nu}_{\ \ \alpha} \td{X}^\alpha$, thus the matrices associated to the representation are given by $[\mathcal{R}(\td{X}^\mu)]_\alpha^{\ \,\beta} = c^{\mu\beta}_{\ \ \ \alpha}$. It is possible to construct another representation from $\mathcal{R}$ via the matrices $\mathcal{R}^*(\td{X}^\mu) = - (\mathcal{R}(\td{X}^\mu))^{\mrm{T}}$, where the superscript $\mrm{T}$ stands for the transpose of the matrix \cite{fuchsbook}. We will call this matrix representation the {\it co-adjoint} representation $\mathcal{R}^*$ of $\mf{an}(n)$. 

In what follows we will work with the co-adjoint representation for $\mf{an}(n)$ since it allows to extend the $(n+1)\times (n+1)$-matrix representation of $\mf{an}(n)$ to include the basis of $\mf{t}$ arranged in an extra column resulting in a $(n+2) \times (n+2)$-matrix representation for the Lie algebra  $\mf{t} \oplus \mf{an}(n)$.  Let us write such matrix representation explicitly. The basis for $\mf{an}(n)$ is given by $n+1$ matrices of size $(n+1) \times (n+1)$. The basis of the Lie algebra $\mf{t} \oplus \mf{an}(n)$ can be represented in terms of $(n+2)\times(n+2)$-matrices. The matrices corresponding to $\mf{t}$ read
	\beq \label{tan-matbasis}
	P_\mu = 
	\begin{pmatrix}
	    0_{(n+1) \times (n+1)}           &&  \mathbf{u}_\mu     \\
	         &&                             &&                    \\
	     \mathbf{0}_{(n+1)}^{\, \mrm{T}}  &&    0             
	\end{pmatrix},
	\eeq
where $\mathbf{0}_n$ and $\mathbf{0}_{(n+1)}$ are $n$ and $(n+1)$-component zero vectors, respectively, and $\mathbf{u}_\mu = (0,\ldots, 1,\ldots, 0)$ is a $(n+1)$-component vector with $1$ in the $\mu$th-entry for $\mu = 0,\ldots, n$. The matrices representing the $\mf{an}(n)$ sector are given by
	\beq \label{tan-matbasis1}
	\td{X}^0 = -\frac{1}{\kappa}
	\begin{pmatrix}
	                       && \mathbf{0}_n^{\, \mrm{T}}          &&   &&              		 \\
	\mathbf{0}_{(n+2)}     && \mathbbm{1}_{n \times n}  			&&   && \mathbf{0}_{(n+2)}    \\
	                       && \mathbf{0}_n^{\, \mrm{T}}          &&   &&                            
	\end{pmatrix}
	\quad \text{and} \quad
	\td{X}^a = \frac{1}{\kappa}
	\begin{pmatrix}
	              		&&  \mathbf{e}_a^{\, \mrm{T}}    &&              \\
	 \mathbf{0}_{(n+2)}  &&    0_{n \times n}    &&    	 \mathbf{0}_{(n+2)}		 \\
	             		&&             \mathbf{0}_n^{\, \mrm{T}}            		&&              
	  \end{pmatrix},
	\eeq
where $\mathbf{e}_a = (0,\ldots, 1,\ldots, 0)$ is a $n$-component vector with $1$ in the $a$th-entry for $a= 1,\ldots, n$. A general group element $d \in \td{D}(T) = T \times AN(n)$ can be expressed as $d = t\ g$ where $t$ is a pure translation and $g$ is a pure $AN(n)$ element. The explicit matrix form of the group element is given by
	\beq \label{decomp}
	d = 
	\begin{pmatrix}
	        \td{g} &&  \mathbf{x}_{n+1} \\
	               &&                   \\
	       \mathbf{0}^{\, \mrm{T}}_{n+1} && 1
	\end{pmatrix},  \quad \text{with} \quad 
	g = 
	\begin{pmatrix}
	\td{g} && \mathbf{0}_{n+1} \\
	\mathbf{0}^{\, \mrm{T}}_{n+1} && 1
	\end{pmatrix}, \quad 
	t = 
	\begin{pmatrix}
	         \mathbbm{1}_{(n+1) \times (n+1)} && \mathbf{x}_{n+1} \\
	                                          &&         \\
	        \mathbf{0}_{n+1}^{\, \mrm{T}}          && 1
	\end{pmatrix},   
	\eeq
where $\td{g} \in AN(n)$ is a $(n+1) \times (n+1)$ matrix and $\mathbf{x}_{n+1} = (x^0, x^a)$ is a $(n+1)$-component vector with real entries $x^\mu \in \mathbb{R}$ that parametrize the group elements of $T \sim \mathbb{R}^{n,1}$, $t = e^{x^\mu P_\mu}$. The $AN(n)$ group can be parametrized in different ways using a set of real coordinates $\{p_0, \ldots, p_n\}$ so the matrix group entries are $\td{g}_{ij} (p)$. 

Before presenting the Poisson brakets for particular parametrizations we will first write down the results in a general, coordinate independent, form.  Using the matrix representation for $\mathcal{D}(\mf{t})$ the Poisson brackets for the Heisenberg double are determined by 
	\beq  \label{rmatrix4}
	\{d\, \overset{\otimes}{,}\, d\}  = - [r, d \otimes d]_+,
	\eeq
with the $r$-matrix given by \eqref{rmatrixan}. In order to see how the Poisson brackets can be read off the expression above, one should recall that, in a simplified case in which $d$ is a $2\times 2$-matrix, the left hand side of \eqref{rmatrix4} would be given, for example, by
	\beq  \label{lhsexpl}
	\{d\, \overset{\otimes}{,}\, d\} = 
	\begin{pmatrix}
	\{d_{11}, d_{11}\} && \{d_{11}, d_{12}\} && \{d_{12}, d_{11}\} && \{d_{12}, d_{12}\}  \\
	\{d_{11}, d_{21}\} && \{d_{11}, d_{22}\} && \{d_{12}, d_{21}\} && \{d_{12}, d_{22}\}  \\
	\{d_{21}, d_{11}\} && \{d_{21}, d_{12}\} && \{d_{22}, d_{11}\} && \{d_{22}, d_{12}\}  \\
	\{d_{21}, d_{21}\} && \{d_{21}, d_{22}\} && \{d_{22}, d_{21}\} && \{d_{22}, d_{22}\}
	\end{pmatrix},
	\eeq
whereas the components in the right hand side of \eqref{rmatrix4} are simply those of the product of the matrices $- \left( r (d \otimes d) - (d \otimes d) r \right)$. An explicit calculation of \eqref{rmatrix4} leads thus to the following Poisson brackets 
	\beq  \label{genpbanxx}
	\{x^0, x^a\} = \frac{1}{\kappa}\, x^a \qquad \text{and} \qquad \{x^a, x^b\} = 0,
	\eeq
	\beq  \label{genpbanpp}
	\{g_{ij}(p), g_{kl}(p)\} = 0,
	\eeq
and
	\beq  \label{genpbanxp}
	\{x^\mu, g\} = - \tilde{X}^\mu\ g,
	\eeq
where $g_{ij}(p)$ are the entries of the matrix representing a pure $AN(n)$ element in $T \times AN(n)$ and the last bracket can be written explicitly in terms of coordinates $x^\mu$ and the entries of $g$ as $\{x^\mu, g_{ij}(p)\} = [\tilde{X}^\mu\ g]_{ij}(p)$ i.e. using the explicit parametrization of the matrix representation of $g$ \footnote{It is worth to mention that using the adjoint representation $\mathcal{R}$ for $\mf{an}(n)$, we can describe the right decomposition of $d = g\ t \in T \times AN(n)$. The Poisson brackets are again obtained from \eqref{rmatrix4} just changing the sign of the $r$-matrix, $r \to -r$ and are given by \eqref{genpbanxx}, \eqref{genpbanpp} and
	\beq  \label{genpbanrightxp}
	\{x^\mu, g\} = g\ \tilde{X}^\mu.
	\eeq}
From \eqref{genpbanpp} we can see that for any coordinates for the momentum group manifold the Poisson brackets are
	\beq \label{genpbanppbis}
	\{p_\mu, p_\nu\} = 0.
	\eeq

For illustrative purposes we write down the explicit form of the Poisson brackets above for some specific parametrizations widely used in the literature. A pure $AN(n)$ group element $g$ can be written as
	\beq  \label{gbeta}
	g = e^{-\beta p_0 \td{X}^0} e^{-p_1 \td{X}^1 -p_2 \td{X}^2} e^{-(1-\beta) p_0 \td{X^0}},
	\eeq
where the values for $ 0 \leq \beta \leq 1$ describe the different coordinates systems. Among them, $\beta = 0$ corresponds to the ``time-to-the-right" parametrization and in this case $\{p_\mu\}$ are known as {\it bicrossproduct coordinates} (since they are associated with the so-called bicrossproduct basis of the $\kappa$-Poincar\'e algebra  \cite{Majid:1994cy}), $\beta = 1$ corresponds to the time-to-the-left parametrization and $\beta = \tfrac{1}{2}$ to the time-symmetric parametrization \cite{Agostini:2003vg}. The general group element \eqref{gbeta} gives rise to the following general element $d = t\ g \in T \times AN(n)$
	\beq  \label{dmatrix}
	d = 
	\begin{pmatrix}
	1 			&& -\frac{1}{\kappa} e^{\frac{(1-\beta)p_0}{\kappa}} \mathbf{p}_n^{\mrm{T}} &&   \\
	\mathbf{0}_{n} && \mathrm{e}^{\frac{p_0}{\kappa}} \mathbbm{1}_{n \times n} && \mathbf{x}_{(n+1)} \\
	    &&  \mathbf{0}_{n}^{\mrm{T}} && 1
	\end{pmatrix},
	\eeq
where $\mathbf{p}_n = ( p_1,\ldots,p_n)$ and $\mathbf{x}_{(n+1)} = (x^0, x^1,\ldots,x^n)$. Notice that the $AN(n)$ part is an upper-triangular matrix and this is a consequence of choosing the co-adjoint representation for the basis of its Lie algebra. From \eqref{genpbanxp} and using that $\{x^\mu,f(p)\} = \{x^\mu,p_\nu\} \tfrac{\partial f(p)}{\partial p_\nu}$ we find that the Poisson brackets for the different coordinates systems labelled by $\beta$ are
	\begin{align}      \label{Poissonbraxp}
	\{x^0,p_0\} &= 1,  \nonumber  \\
	\{x^0, p_a\} &= -\frac{p_a}{\kappa}(1-\beta), \nonumber \\
	\{x^a,p_0\} &= 0, \nonumber \\
	\{x^a, p_b\} &= \delta^a_b\, \e^{\frac{p_0}{\kappa}\beta}.
	\end{align}
It is worth to write down the relations for the cases $\beta = 0, \tfrac{1}{2}, 1$ and their first order expansions in $\tfrac{1}{\kappa}$. For the time-to-the-right case $\beta = 0$ the deformed brackets at {\it all orders} in $\kappa$ read
\beq      \label{Poissonrightan}
	\{x^0,p_0\} = 1\,,\qquad 
	\{x^0, p_a\} = -\frac{p_a}{\kappa}\,,\qquad 	\{x^a,p_0\} = 0\,,\qquad
	\{x^a, p_b\} = \delta^a_b\ .
	\eeq
The case $\beta = 1$ or time to the left has the following Poisson brackets
	\beq     \label{Poissonleftan}
	\{x^0,p_0\} = 1\,,\qquad   
	\{x^0, p_a\} = 0\,,\qquad 
	\{x^a,p_0\} = 0\,,\qquad 
	\{x^a, p_b\} = \delta^a_b\, \e^{\frac{p_0}{\kappa}},
	\eeq
and to the second order in $\tfrac{1}{\kappa}$ the only relevant relation is 
	\beq      \label{Poissonleftan1order}
	\{x^a, p_b\} = \delta^a_b\, \left( 1 + \frac{p_0}{\kappa} + \frac{p_0^2}{2 \kappa^2} + \mathcal{O}\left(\frac{1}{\kappa^3}\right) \right).
	\eeq
Finally, the time symmetric case $\beta = \tfrac{1}{2}$ gives
	\beq     \label{Poissontsymm}
	\{x^0,p_0\} = 1\,,\qquad 
	\{x^0, p_a\} = -\frac{p_a}{2 \kappa}\,,\qquad 
	\{x^a,p_0\} = 0\,,\qquad 
	\{x^a, p_b\} = \delta^a_b\, \e^{\frac{p_0}{2 \kappa}},
	\eeq
for which the expansion up to second order is
	\beq  \label{Poissontsymm1order}
	\{x^a, p_b\} = \delta^a_b\, \left( 1 + \frac{p_0}{2 \kappa} + \frac{p_0^2}{8 \kappa^2} + \mathcal{O}\left(\frac{1}{\kappa^3}\right) \right).
	\eeq
The brackets for the time-to-the-right parametrization \eqref{Poissonrightan}, $\beta = 0$ coincide with those found in \cite{Lukierski:1993wx, Arzano:2010kz, AmelinoCamelia:1997jx, AmelinoCamelia:2011nt}. \footnote{In \cite{AmelinoCamelia:2011nt} the Poisson brackets are $\{x^0, x^a\} = -\frac{1}{\kappa} x^a$ and $\{x^0,p_a\} = \frac{p_a}{\kappa}$. This difference in the sign can be traced back to a matter of convention on the signature of the embedding Minkowski space for the $AN(n)$ manifold.}. It is worth mentioning that in \cite{Arzano:2010kz, AmelinoCamelia:2011nt} the Poisson brackets were obtained using a different procedure starting from the Kirillov symplectic form \cite{kirillov1976} to write down the kinetic term for the reduced action of the relativistic particle. \footnote{For a different approach to deformed phase spaces which makes use of the theory of Hopf algebroids see \cite{Meljanac:2014jsa, Lukierski:2015zqa, Lukierski:2016utz}} In our approach the Poisson structure is derived purely in terms of the algebraic structure of the generators of the momentum group manifold.

Before closing the section a remark is in order. We determined a symplectic Poisson structure on the phase space group manifold $\Gamma=T \times AN(n)$ using the Heisenberg double construction. The alternative Drinfeld double construction, has the property of being compatible with the Lie group multiplication and indeed is related to the symmetries of the phase space \cite{Bonzom:2014wva}. Using \eqref{rmatrix3} with the commutator, i.e. $\{d\, \overset{\otimes}{,}\, d\}  = - [r, d \otimes d]_-$, we can find the Poisson brackets associated to the Drinfeld double structure. These are given again by \eqref{genpbanxp}, \eqref{genpbanpp} but now $\{x^\mu, g\} = 0$ for all $\mu$, so the ``cross" brackets between position and momenta vanish identically.

\section{Deforming momentum space to the group $SL(2,\mathbb{R})$}
\label{sec:sl2}

We now discuss another important example of group valued momenta which emerge in the context of three-dimensional gravity and, like the $AN(n)$ case discussed above, are associated to a deformation of the Poincar\'e group.

As it is well known, in three space-time dimensions gravity does not possess local degrees of freedom. Point-like particles can be included in theory as {\it topological defects}. For instance, coupling a spinless particle at rest to gravity results in a conical metric with  the particle sitting at the tip of the cone \cite{thooft1984}. This conical metric can be pictured as a wedge cut out from the spatial plane characterized by a deficit angle proportional to the mass of the particle $\alpha = 8\pi G m$, where $G$ is Newton's gravitational constant which in two dimensions has units of inverse mass. The deficit angle is described through a rotation by $8\pi Gm$ which is captured by calculating the holonomy of the flat connection around the location of the particle. Thus the momentum at rest of the particle is determined by a rotation proportional to $m G$ i.e. a group element belonging to $SL(2,\mathbb{R})$, the (double cover) of the group of isometries of three-dimensional Minkowski space. A description of a moving defect can be obtained by boosting the conical metric, in this case the three momentum of the particle will be a general element of $SL(2,\mathbb{R})$ \cite{Matschull:1997du, lotito2014}. Various treatments exist for the description of the phase space of point particles coupled to gravity in three dimensions \cite{Matschull:1997du, Meusburger:2003ta, Osei:2011ig} and its symmetries \cite{Ballesteros:2010zq, Ballesteros:2013dca}. 

Here we will show how our general prescription for defining Poisson structures on group manifold phase spaces can be applied to this case where the phase space is given by the cartesian product of the group of translations times the double cover of the three dimensional Poincar\'e group, $\mathbb{R}^{2,1} \times SL(2,\mathbb{R})$. In particular we take as the only input of our construction the algebraic properties determined by the Lie group structure of the (extended) momentum space $SL(2,\mathbb{R})$.

As in the previous Section, we start from the Lie algebra $\mf{t} \oplus \mf{sl}(2,\mathbb{R})$ and the infinitesimal counterpart of the Poisson bivector, the co-commutator $\delta_{\mf{t} \oplus \mf{sl}(2,\mathbb{R})}$. Let us denote with $\{P_\mu\}$ and $\{\tilde{X}^\mu\}$ the generators of $\mf{t}$ and $\mf{sl}(2,\mathbb{R})$, respectively, whose Lie algebras are determined by the following Lie brackets
	\beq  \label{lietsl2r}
	[P_\mu, P_\nu] = 0, 
	\eeq
for all $\mu = 0,1,2$ and
	\beq  \label{lietsl2r2}
	[\tilde{X}^0, \tilde{X}^1] = - \frac{1}{\ell} \tilde{X}^2, \quad [\tilde{X}^0, \tilde{X}^2] = \frac{1}{\ell} \tilde{X}^1, \quad [\tilde{X}^1, \tilde{X}^2] = \frac{1}{\ell} \tilde{X}^0.
	\eeq
The Lie brackets of $\mf{sl}(2,\mathbb{R})$ are obtained from the usual relation $[\tilde{X}_{\mu}, \tilde{X}_{\nu}] = \frac{1}{\ell} \epsilon_{\mu\nu\sigma}  \tilde{X}^{\sigma}$, raising and lowering indices using a ``mostly minus" Minkowski metric and the totally skew-symmetric Levi-Civita pseudotensor is such that $\epsilon_{012} = 1$ \cite{schroerswilhelm2014}. 
The direct sum $\mf{t} \oplus \mf{sl}(2,\mathbb{R})$ can be made into a Lie algebra extending the inner product as in \eqref{innerprodD} and defining Lie brackets such that the product is ad-invariant. The resulting Lie algebra structure is given by the brackets (\ref{lietsl2r}, \ref{lietsl2r2}) together with
	\beq  \label{tplussl2rlie}
	\begin{split}
	[P_0, \tilde{X}^1] = \frac{1}{\ell} P_2, \quad [P_0, \tilde{X}^2] = -\frac{1}{\ell} P_1, \quad [P_1, \tilde{X}^0] = \frac{1}{\ell} P_2, \\
	 [P_1, \tilde{X}^2] = - \frac{1}{\ell} P_0, \quad [P_2, \tilde{X}^0] = - \frac{1}{\ell} P_1, \quad [P_2, \tilde{X}^1]= \frac{1}{\ell} P_0.  
	 \end{split}
	\eeq
We can now introduce co-commutators on $\mf{t} \oplus \mf{sl}(2,\mathbb{R})$ using an $r$-matrix analogous to \eqref{rmatrixan}. Denoting the complete set of generators $Z_A = \{P_\mu, \tilde{X}^\mu\}$ we have for the co-commutator the explicit relations
	\beq  \label{inclusionsl2r}
	\delta_{\mf{t} \oplus \mf{sl}(2, \mathbb{R})} (\tilde{X}^\mu) = 0,
	\eeq
	\beq   \label{inclusionsl2r2}
	\delta_{\mf{t} \oplus \mf{sl}(2, \mathbb{R})} (P_0) = \frac{2}{\ell}\, P_1 \wedge P_2, \quad \delta_{\mf{t} \oplus \mf{sl}(2, \mathbb{R})} (P_1) = \frac{2}{\ell}\, P_0 \wedge P_2, \quad \delta_{\mf{t} \oplus \mf{sl}(2, \mathbb{R})} (P_2) = -\frac{2}{\ell}\, P_0 \wedge P_1\,.
	\eeq
 These co-commutators turn $\mf{t} \oplus \mf{sl}(2,\mathbb{R})$ into a Lie bi-algebra i.e. a {\it classical double} which we denote as $\td{\mathcal{D}}(\mf{t})$. 
We can represent the generators as matrices using the co-adjoint representation, as in the previous Section. In such representation the generators $Z_A$ can be written as $4\times 4$-matrices as
	\beq  \label{slP0}
	P_\mu =	
	\begin{pmatrix}
	    0_{3 \times 3} &&  \mathbf{u}_\mu \\
	\mathbf{0}^{\mrm{T}}_3 && 0
	\end{pmatrix},
	\eeq	
where $\mathbf{u}_\mu$ is a $3$-component vector with 1 in the $\mu$-entry for $\mu = 0, \ldots, 2$ and as
	\beq \label{slXs}
	\tilde{X}^0 = \frac{1}{\ell}
	\begin{pmatrix}
	0 && 0 && 0 && 0 \\
	0 && 0 && 1 && 0 \\
	0 && -1 && 0 && 0 \\
	0 && 0 && 0 && 0
	\end{pmatrix},
	\quad
	\tilde{X}^1 = \frac{1}{\ell}
	\begin{pmatrix}
	0 && 0 && -1 && 0 \\
	0 && 0 && 0 && 0 \\
	-1 && 0 && 0 && 0 \\
	0 && 0 && 0 && 0
	\end{pmatrix},
	\quad
	\tilde{X}^2 = \frac{1}{\ell}
	\begin{pmatrix}
	0 && 1 && 0 && 0 \\
	1 && 0 && 0 && 0 \\
	0 && 0 && 0 && 0 \\
	0 && 0 && 0 && 0
	\end{pmatrix}.
	\eeq
A general element of the $T \times SL(2,\mathbb{R})$ group manifold can be decomposed in terms of a pure translation and a pure $SL(2,\mathbb{R})$ transformation as $d = t g$ with the following general matrix representation
	\beq \label{decompsl2r}
	d = 
	\begin{pmatrix}
	       \td{g} &&  \mathbf{x}_{n+1} \\
	       \mathbf{0}^{\, \mrm{T}}_{n+1} && 1
	\end{pmatrix},  
	\eeq
where $\tilde{g} \in SL(2,\mathbb{R})$ is a $3 \times 3$ matrix and
	\beq  \label{gandtsl2r}
	g = 
	\begin{pmatrix}
	\td{g} && \mathbf{0}_{n+1} \\
	\mathbf{0}^{\, \mrm{T}}_{n+1} && 1
	\end{pmatrix}, \quad  \text{and} \quad
	t = 
	\begin{pmatrix}
	\mathbbm{1}_{(n+1) \times (n+1)} && \mathbf{x}_{n+1} \\
	\mathbf{0}_{n+1}^{\, \mrm{T}}          && 1
	\end{pmatrix},   
	\eeq
are the matrices representing a pure Lorentz transformation and a pure translation, respectively. 

As in the previous section we can now use the Heisenberg double relation $\{d\, \overset{\otimes}{,}\, d\}  = - [r, d \otimes d]_+$ to write derive the Poisson brackets for the $T \times SL(2,\mathbb{R})$ phase space. The general expressions for the Poisson bracket in terms of the coordinates $x^{\mu}$ appearing in \eqref{decompsl2r} and of momenta are given by
	\beq  \label{genpbsl2rxx}
	\{x^0, x^1\} = -\frac{1}{\ell}\, x^2 \qquad, \{x^0, x^2\} = \frac{1}{\ell}\, x^1 \qquad \text{and} \qquad \{x^1, x^2\} = \frac{1}{\ell}\, x^0,
	\eeq
	\beq  \label{genpbsl2rpp}
	\{g_{ij}(p), g_{kl}(p)\} = 0  \implies \{p_\mu, p_\nu\} = 0,
	\eeq
 and
	\beq  \label{genpbsl2rxp}
	\{x^\mu, g\} = - \tilde{X}^\mu\, g.
	\eeq
Using the adjoint representation, instead of the co-adjoint one, the Poisson brackets are given again by \eqref{genpbsl2rxx} and \eqref{genpbsl2rpp} but now the mixed brackets read $\{x^\mu, g\} = - g\, \td{X}^\mu$. These brackets coincide with those found in \cite{Matschull:1997du}, one of the earliest descriptions of the phase space of gravitating particle in three dimensions based on the Hamiltonian treatment of the reduced action of the gravity-plus-particle system. 
 
It is instructive to focus on a given parametrization of the momentum group manifold. We consider the ``exponential coordinates" \cite{lotito2014} for which $g \in SL(2, \mathbb{R})$ is obtained as $g = e^{-p_\mu \td{X}^\mu}$, $\mu= 0,1,2$. The mass parameter $\ell = 1/4\pi G$ is determined by the three-dimensional Newton's constant  \cite{thooft1984} and in the limit $G\rightarrow 0$ one recovers the usual flat momentum space $\mathbb{R}^{2,1}$. The matrix that describes the general group element $d \in T \times SL(2,\mathbb{R})$ is given as
	\beq  \label{gsl2r}
	d = 
	\begin{pmatrix}
	\frac{p_0^2 - (p_1^2 + p_2^2) \cos \frac{p}{\ell}}{p^2} && \frac{p_0 p_1 - p_0 p_1 \cos \frac{p}{\ell} - p_2 p \sin \frac{p}{\ell}}{p^2} && \frac{p_0 p_2 - p_0 p_2 \cos \frac{p}{\ell} + p_1 p \sin \frac{p}{\ell}}{p^2}  && x^0 \\
	-\frac{p_0 p_1 - p_0 p_1 \cos \frac{p}{\ell} + p_2 p \sin \frac{p}{\ell}}{p^2} && \frac{-p_1^2 + (p_0^2 - p_2^2) \cos \frac{p}{\ell}}{p^2} && -\frac{p_1 p_2 - p_1 p_2 \cos \frac{p}{\ell} - p_0 p \sin \frac{p}{\ell}}{p^2} && x^1 \\
	-\frac{p_0 p_2 - p_0 p_2 \cos \frac{p}{\ell} - p_1 p \sin \frac{p}{\ell}}{p^2} && -\frac{p_1 p_2 - p_1 p_2 \cos \frac{p}{\ell} - p_0 p \sin \frac{p}{\ell}}{p^2} && \frac{-p_2^2 + (p_0^2 - p_1^2) \cos \frac{p}{\ell}}{p^2} && x^2 \\
	0 && 0 && 0 && 1 
	\end{pmatrix},
	\eeq
where $p^2 = p_0^2 - p_1^2 - p_2^2$. The explicit, all order, relations for \eqref{genpbsl2rxp} in terms of the coordinates for the group $(x^\mu, p_\mu)$ are rather involved and here we present these Poisson brackets at first order in the deformation parameter $\frac{1}{\ell}$
	\begin{align} \label{pbsl2xp}
	\{x^0, p_0\} &= 1, \qquad \;\;\,\  \{x^0, p_1\} = -\frac{1}{\ell}\frac{p_2}{2}, \quad  \{x^0, p_2\} = \frac{1}{\ell} \frac{p_1}{2}, \nonumber \\
	\{x^1, p_0\} &= -\frac{1}{\ell}\frac{p_2}{2}, \quad \{x^1, p_1\} = 1, \qquad \quad \{x^1, p_2\} =- \frac{1}{\ell} \frac{p_0}{2}, \nonumber \\
	\{x^2, p_0\} &= \frac{1}{\ell}\frac{p_1}{2}, \quad \;\;\;  \{x^2, p_1\} = \frac{1}{\ell} \frac{p_0}{2}, \quad \;\ \{x^2, p_2\} = 1 .
	\end{align}
These relations can be written in a compact way as 
	\beq \label{pbsl2compact}
	\{x_\mu, p_\nu\} = \eta_{\mu\nu} + \frac{1}{\ell}\, \epsilon_{\mu\nu\alpha}\, \frac{p^\alpha}{2}.
	\eeq

We can consider cartesian coordinates on the group manifold by transforming the general group element in exponential coordinates \eqref{gsl2r} through the relations \cite{lotito2014} 
	\beq \label{expovscart}
	\td{p}_\mu = \frac{\sin \frac{p}{\ell}}{p} p_\mu,
	\eeq 
We find that the group element parametrized by cartesian coordinate is represented by the following matrix	
	\beq \label{dcart}
	d = 
	\begin{pmatrix}
	\frac{\td{p}_0^2 - (\td{p}_1^2 + \td{p}_2^2) \sqrt{1 - \frac{\td{p}^2}{\ell^2}}}{\td{p}^2} && \frac{\td{p}_0 \td{p}_1}{\left(1 + \sqrt{1 - \frac{\td{p}^2}{\ell^2}}\right)\ell^2} - \frac{\td{p}_2}{\ell}  && \frac{\td{p}_0 \td{p}_2}{\left(1 + \sqrt{1 - \frac{\td{p}^2}{\ell^2}}\right)\ell^2} + \frac{\td{p}_1}{\ell} && x^0 \\
	-\frac{\td{p}_0 \td{p}_1}{\left(1 + \sqrt{1 - \frac{\td{p}^2}{\ell^2}}\right)\ell^2} - \frac{\td{p}_2}{\ell} && -\frac{\td{p}_1^2 - (\td{p}_0^2 - \td{p}_2^2) \sqrt{1 - \frac{\td{p}^2}{\ell^2}}}{\td{p}^2} && -\frac{\td{p}_1 \td{p}_2}{\left(1 + \sqrt{1 - \frac{\td{p}^2}{\ell^2}}\right)\ell^2} - \frac{\td{p}_0}{\ell} && x^1 \\
	-\frac{\td{p}_0 \td{p}_2}{\left(1 + \sqrt{1 - \frac{\td{p}^2}{\ell^2}}\right)\ell^2} + \frac{\td{p}_1}{\ell} && -\frac{\td{p}_1 \td{p}_2}{\left(1 + \sqrt{1 - \frac{\td{p}^2}{\ell^2}}\right)\ell^2} + \frac{\td{p}_0}{\ell} && -\frac{\td{p}_2^2 - (\td{p}_0^2 - \td{p}_1^2) \sqrt{1 - \frac{\td{p}^2}{\ell^2}}}{\td{p}^2} && x^2 \\
	0 && 0 && 0 && 1 
	\end{pmatrix}
	\eeq
where $\td{p}^2 = \td{p}_0^2 - \td{p}_1^2 - \td{p}_2^2$. The Poisson brackets up to first order have the same form as in equations \eqref{genpbsl2rpp}, \eqref{genpbsl2rxx} and \eqref{pbsl2xp}, that is, in compact form we have $\{x_\mu, \td{p}_\nu\} = \eta_{\mu\nu} + \frac{1}{\ell}\, \epsilon_{\mu\nu\alpha}\, \frac{\td{p}^\alpha}{2}$. These Poisson brackets coincide at first order in the deformation parameter with those found in \cite{bianco2013}. The Poisson brackets in \cite{bianco2013} are given by $\{x^\mu, x^\nu\} = \ell \epsilon^{\mu\nu}_{\;\;\;\, \rho}\, x^\rho$, $\{p_\mu, p_\nu\} = 0$ and $\{p_\mu, x^\nu\} \simeq -\delta^\nu_\mu + \frac{\ell}{2}\ \epsilon_{\mu}^{\;\ \nu\rho}\ p_\rho$, where indices are raised and lowered with a mostly plus Minkowski metric and $\epsilon_{\mu\nu\alpha}$ is such that $\epsilon_{012} = -1$. For comparison with our results we must consider that in \cite{bianco2013} the Lie algebra $\mf{sl}(2,\mathbb{R})$ is defined by the commutators $[X^\mu, X^\nu] = \epsilon^{\mu\nu}_{\;\;\;\, \rho} X^\rho$, where $\{X^\mu\}$ is the basis for $\mf{sl}(2, \mathbb{R})$, while the convention we are following is that of \cite{schroerswilhelm2014}, where $[X_\mu, X_\nu] = \epsilon_{\mu\nu\sigma}X^\sigma$ with a mostly minus Minkowski metric for lowering and raising indices and $\epsilon_{012} = 1$. Therefore, in order to compare we must identify $X^1 \rightarrow -X^2$, $X^2 \rightarrow -X^1$, which translates into the following identifications for the phase space coordinates $x^1 \rightarrow -x^2$, $x^2 \rightarrow -x^1$, $p^1 \rightarrow -p^2$ and $p^2 \rightarrow -p^1$. Taking into account these identifications we can compare and we indeed get the same results up to first order in the deformation parameter as in \cite{bianco2013}.


\section{Composite systems of classical particles}
\label{sec:composite}

The treatment of multi-particle systems with group valued momenta has notoriously been controversial. Indeed when considering the deformations of translations associated to such models it is often assumed that {\it any} elementary system, quantum or classical, exhibits a non-trivial composition of momenta associated with the non-abelian group multiplication of momentum space. This leads to blatant contradictions with the known laws of kinematics of macroscopic bodies known in the literature as ``the soccer ball problem" \cite{amelinosoccer2011, hossenfelder2014}.
In this Section we discuss the composition of momenta in classical systems both in undeformed relativistic kinematics and in the deformed case discussed so far.
We will show that under the assumption that phase spaces of composite systems with group valued momenta are given by the {\it cartesian product} of their components the total momentum of a multi-particle system is given by the abelian sum of the individual momenta of the components. We briefly mention how, upon quantization the non-abelian structure of momentum space comes into play and momenta associated to multiparticle states must compose according to a non-abelian composition rule. Thus as long as we do not observe ``quantum soccer balls" in experiments there is no obvious problem with the composition of momenta in systems with a Lie group momentum space.

Let us start by recalling that, as seen in the previous Sections, the phase space of a classical relativistic (spinless) point particle is just a direct sum vector space given by $\Gamma = \mathbb{R}^{3,1} \oplus \mathbb{R}^{3,1}{}^*$. Let us restrict to momentum space $\Gamma_p = \mathbb{R}^{3,1}{}^*$. The dual space to Minkowski space is isomorphic to Minkowski space itself as vector space $\mathbb{R}^{3,1}{}^*\simeq \mathbb{R}^{3,1}$ equipped with the usual (abelian) vector addition $p+q$ for $p,q \in \Gamma_p$. Let us recall that an {\it observable} $\mathcal{O}$ for a classical system is a map $\mathcal{O}:\Gamma \rightarrow \mathbb{R}$. A particular set of observables is given by the components of momentum $\mathcal{P}^{\mu}$, so that to $p\in \Gamma_p$ one associates a four-vector $\mathcal{P}^{\mu}(p)=p^{\mu}\in \mathbb{R}^{3,1}$. 

Now let us consider a composite systems made of two particles. The phase space of such system will be given\footnote{See \cite{Geroch:1985ci} pag. 184 and for a more ``philosophical" discussion see \cite{Aerts:1978}.} by the cartesian product of the respective phase spaces $\Gamma \equiv \Gamma_1 \times \Gamma_2 \simeq  \Gamma_1 \oplus \Gamma_2$ where the last isomorphism holds if $\Gamma_1$ and $\Gamma_2$ are \emph{vector spaces}. Again we can focus on the momentum sector of phase space which will be given by $\Gamma_p=\Gamma_{p1} \oplus \Gamma_{p2}$. Given a point $(p_1,p_2)\in \Gamma_{p1} \oplus \Gamma_{p2} \simeq \mathbb{R}^{3,1} \oplus \mathbb{R}^{3,1}$ we want to see how the single particle momenta combine to give the {\it total} momentum associated to such point in phase space. In order to do so let us first recall how the vector composition is defined for direct sums of vector spaces. Given $(p_1,p_2), (p'_1,p'_2) \in \Gamma_{p1} \oplus \Gamma_{p2}$ we can extend the vector addition $+$ defined in $\Gamma_p$ to $\Gamma_{p1} \oplus \Gamma_{p2}$ as follows
\beq\label{cartadd1}
(p_1,p_2) + (p'_1,p'_2) \equiv (p_1+p'_1, p_2+ p'_2)\,.
\eeq
An observation which will be crucial for what follows is that, given the composition above, any element $(p_1,p_2)\in \Gamma_{p1} \oplus \Gamma_{p2}$ can be written as
\beq\label{cartsum1}
(p_1,p_2) = (p_1, 0) + (0 ,p_2)\,. 
\eeq

Now let us look at the observable $\mathcal{P}^{\mu}(p_1,p_2)$ associated to the momentum space point $(p_1,p_2) \in \Gamma_{p1} \oplus \Gamma_{p2}$. Starting from the definition of (coordinate) functions on the cartesian product of spaces we have
\beq
\mathcal{P}^{\mu}(p_1,p_2) \equiv (\mathcal{P}^{\mu}(p_1),\mathcal{P}^{\mu}(p_2)) = (p^{\mu}_1,p^{\mu}_2) = (p^{\mu}_1, 0) + (0 ,p^{\mu}_2) = \mathcal{P}^{\mu}(p_1,0) + \mathcal{P}^{\mu}(0,p_2)
\eeq
where in the fourth term we used the analogous of \eqref{cartsum1} for $\mathbb{R}^{3,1} \oplus \mathbb{R}^{3,1}$. 

Now let us consider for example $\mathcal{P}^{\mu}(p_1,0)$, this is the momentum observable associated to the two-particle system when particle 2 has vanishing momentum and thus we can make the identification $ \mathcal{P}^{\mu}(p_1,0) = (p^{\mu}_1, 0) \rightarrow p^{\mu}_{1}$. Mathematically this is reflected in the fact that for a generic group $G$ the isomorphism $G \times \{ \mathbbm{1}\}\simeq G$, where $\mathbbm{1}$ is the identity element, holds and, in particular in our case $\mathbb{R}^{3,1} \times \{0\}\simeq \mathbb{R}^{3,1}$. Such isomorphism maps the sum $+$ \emph{restricted} to  $ (\mathbb{R}^{3,1} \times \{0\}) \times ( \{0\} \times  \mathbb{R}^{3,1})$ to the ordinary sum defined on $\mathbb{R}^{3,1} \times \mathbb{R}^{3,1}$. Thus the {\it total momentum} of the system is the map that associates, via the isomorphsims above, to the observable  $\mathcal{P}^{\mu}(p_1,p_2) = \mathcal{P}^{\mu}(p_1,0) + \mathcal{P}^{\mu}(0,p_2)$ the four-vector $p^{\mu}_{12}$ given by
\beq
\mathcal{P}^{\mu}(p_1,p_2) \rightarrow p^{\mu}_{12} = p^{\mu}_1 + p^{\mu}_2\,,
\eeq
obtained from the vector sum of the single particle four-momenta $p^{\mu}_1, p^{\mu}_2$. To summarize: starting from basic first principles we reproduced the familiar composition of momenta we are accustomed to. Even though the discussion above might seem redundant, it helps clarifying the subtleties one faces in treating group valued momenta as we are now going to see.

Let us now consider the case of classical deformed kinematics in which the momentum (vector) space is replaced by a group manifold. As we have seen in the previous sections the deformed phase space will be given by $\Gamma = \mathbb{R}^{3,1} \times G$ where $G$ is a four dimensional non-abelian Lie group. 
In analogy with the undeformed case, observables will be given by maps from $\Gamma $ to the real numbers $\mathbb{R}$. In particular {\it a} four-momentum observable $ \mathcal{P}^{\mu}$ associates a four-vector to the any momentum space element $\pi \in \Gamma_{\pi} = G$ namely: $\mathcal{P}^{\mu}(\pi) = \pi^{\mu}$.

In complete analogy with the undeformed case, the phase space of a composite system of two particles is given by $\Gamma = \Gamma_1 \times \Gamma_2 = \mathbb{R}^{3,1}_1 \oplus \mathbb{R}^{3,1}_2 \times G_{1} \times G_2$ and the total momentum space is thus $\Gamma^{(2)}_{\pi} = \Gamma_{\pi 1}\times \Gamma_{\pi 2} \equiv G_{1} \times G_2$.

Now let us focus on the single-particle four-momentum observable $\mathcal{P}^{\mu}:  \Gamma_{\pi} \rightarrow \mathbb{R}^{3,1}$. 
This observable corresponds to a choice of coordinate function on the momentum group manifold $G$ \cite{arzano2011}. The first thing to notice is that the Lie group multiplication induces a {\it non-abelian} addition law $\triangleright$ for four-momenta defined  by
\beq
\mathcal{P}^{\mu}(\pi \cdot \tilde{\pi}) \equiv \mathcal{P}^{\mu}(\pi) \triangleright \mathcal{P}^{\mu}(\tilde{\pi}) \,, 
\eeq
for $\pi, \tilde{\pi} \in \Gamma_{\pi} $
The main point we want to stress is that such addition law {\it does not} represent the composition law for classical four-momentum observables. To see that this is the case let us consider the four-momentum observable for the two particle state $(\pi_1,\pi_2)$. From the definition of coordinates on cartesian products of manifolds we have
\beq
\mathcal{P}^{\mu}(\pi_1, \pi_2) \equiv (\mathcal{P}^{\mu}(\pi_1), \mathcal{P}^{\mu}(\pi_2)) = (\pi^{\mu}_1, \pi^{\mu}_2)
\eeq
and from the property of addition on the cartesian product $\mathbb{R}^{3,1}\times\mathbb{R}^{3,1}$
\beq
(\pi^{\mu}_1, \pi^{\mu}_2) = (\pi^{\mu}_1, 0) + (0 , \pi^{\mu}_2) = \mathcal{P}^{\mu}(\pi_1, \mathbbm{1}) + \mathcal{P}^{\mu}(\mathbbm{1}, \pi_2) 
\eeq
where in the last equality we used the fact that $\mathbbm{1}$, the identity element in the group $G$, corresponds to ``vanishing" four-momentum. Using the isomorphism $G \times \{ \mathbbm{1}\} \simeq G$ we can make the obvious identification $\mathcal{P}^{\mu}(\pi_1, \mathbbm{1}) \rightarrow \mathcal{P}^{\mu}(\pi_1) = \pi^{\mu}_1$ and $\mathcal{P}^{\mu}(\mathbbm{1}, \pi_2) \rightarrow \mathcal{P}^{\mu}(\pi_2) = \pi^{\mu}_2$. 

Now let us recall that the group law $\cdot $ can be extended to the cartesian product $G_1\times G_2$ as follows:
\beq
(\pi_1, \pi_2) \cdot (\pi'_1, \pi'_2) \equiv (\pi_1\cdot \pi'_1, \pi_2 \cdot \pi'_2)\,. 
\eeq
In particular {\it any} element $(\pi_1, \pi_2)  \in \Gamma_{\pi 1}\times \Gamma_{\pi 2} \equiv G_{1} \times G_2$ can be written as
\beq
(\pi_1, \pi_2)  \equiv (\pi_1, \mathbbm{1}) \cdot (\mathbbm{1}, \pi_2) \,. 
\eeq
Notice how the \emph{restriction} of the group law $\cdot$ to the cartesian product $(G_1 \times \{ \mathbbm{1}\})\times (\{ \mathbbm{1}\} \times G_2)$ is {\it abelian}. Indeed for the four momentum observable associated to a classical two particle $(\pi_1, \pi_2)$ state we have 
\beq
\mathcal{P}^{\mu}(\pi_1, \pi_2) = \mathcal{P}^{\mu}(\pi_1, \mathbbm{1}) + \mathcal{P}^{\mu}(\mathbbm{1}, \pi_2)\,. 
\eeq
Once such identity is established one can proceed as in the undeformed case and associate, via the identifications $\mathcal{P}^{\mu}(\pi_1, \mathbbm{1}) \rightarrow \mathcal{P}^{\mu}(\pi_1) = \pi^{\mu}_1$ and $\mathcal{P}^{\mu}(\mathbbm{1}, \pi_2) \rightarrow \mathcal{P}^{\mu}(\pi_2) = \pi^{\mu}_2$ a {\it total momentum} four-vector to the two particle momentum observable $\mathcal{P}^{\mu}(\pi_1, \pi_2)$
\beq
\mathcal{P}^{\mu}(\pi_1, \pi_2) \rightarrow \pi_{12}^{\mu} = \pi^{\mu}_1 + \pi^{\mu}_2\,.
\eeq
This shows\footnote{See also \cite{Amelino-Camelia:2014gga} where similar conclusions for a classical system where reached starting from a different perspective.} that for classical systems, which are described by cartesian products, to the observable $\mathcal{P}^{\mu}(\pi_1, \pi_2)$ we can associate a total momentum four vector $\pi_{12}^{\mu} \in \mathbb{R}^{3,1}$ obtained from the {\it ordinary} vector sum of single-particle four-momenta $\pi^{\mu}_1$ and $\pi^{\mu}_2$.\\

The non-abelian Lie group multiplication of momentum space plays however a non-trivial role when we look at the quantum counterparts of our systems. The state of quantum relativistic particle is described by a vector (to be precise a ``ray") in a Hilbert space $|p\rangle \in \mathcal{H}$. An ``one-particle"-observable in this case is a (linear, self-adjoint) operator on $\mathcal{H}$. We focus on four-momentum $P^{\mu}$, i.e. the observable associated to generators of space-time translations. A basis of $\mathcal{H}$ is given by eigenstates of $P^{\mu}$
\beq
P^{\mu} |p\rangle = \mathcal{P}^{\mu}(p) |p\rangle\,.
\eeq
The abstracts kets $|p\rangle $ admit a representation in term of (wave)-functions on Minkowski space, in particular $\langle x| p\rangle = e^{ipx}= e_p(x)$. A two-particle system (we set aside the question of (un)-distinguishability) is described by the tensor product $\mathcal{H}^2 = \mathcal{H}_1 \otimes \mathcal{H}_2$. How do we define the action of the observable $P^{\mu}$ on two-particle states? We can actually derive it using the point-wise multiplication naturally associated to functions on Minkowski space and linearity of observables and, in particular, of four-momenta which act as derivatives on such functions. 
Since for plane waves we have
\beq
e_{p_1} (x) \cdot e_{p_2}(x) \equiv e_{p_1+p_2}(x)
\eeq
by definition 
\beq
P^{\mu}\, (e_{p_1+p_2}(x)) = \mathcal{P}^{\mu} (p_1+p_2)\, e_{p_1+p_2}(x)
\eeq
and for the homomorphism property for coordinate functions on $\mathbb{R}^{3,1}$ we have
\beq
\mathcal{P}^{\mu} (p_1+p_2) =  \mathcal{P}^{\mu} (p_1) + \mathcal{P}^{\mu} (p_2)\,.
\eeq
The main point which should be clearly stressed is that {\it while in the classical case the four-momentum of a two-particle system is given by $\mathcal{P}^{\mu}(p_1, p_2)$ in the quantum domain we have to consider instead $\mathcal{P}^{\mu}(p_1+ p_2)$.}
From the definition of inner product for tensor product states 
\beq
\langle p_1 p_2 | p'_1 p'_2 \rangle = \langle p_1| \otimes \langle p_2|\,\, |p'_1\rangle \otimes |p'_2\rangle \equiv \langle p_1| p'_1\rangle \langle p_2|p'_2\rangle
\eeq
we have
\beq
\mathcal{P}^{\mu} (p_1+p_2) =  \langle p_1 p_2 | P^{\mu} | p_1 p_2 \rangle = \mathcal{P}^{\mu} (p_1) + \mathcal{P}^{\mu} (p_2) = \langle p_1 | P^{\mu} | p_1 \rangle + \langle p_2 | P^{\mu} | p_2 \rangle 
\eeq
from which we can easily derive the action of $P^{\mu}$ on two particle states
\beq
P^{\mu} (|p_1\rangle \otimes |p_2\rangle) = P^{\mu} |p_1\rangle \otimes |p_2\rangle + |p_1\rangle \otimes P^{\mu} |p_2\rangle
\eeq
i.e. such action is dictated by the familiar {\it Leibniz rule}. In the deformed case the story is well known (see e.g. \cite{Arzano:2012bj, Arzano:2013sta}) and we will briefly review the basic concepts referring the reader to \cite{Arzano:2013sta} for further details. When momentum space is represented by a group manifold $G$ basis vectors of the one paricle Hilbert space $\mathcal{H}$ will be given by kets $|\pi \rangle$ labelled by group elements $\pi \in G$. These kets admit a plane wave representation in terms of {\it non-commutative plane waves} $e_{\pi}(x) = \langle x | \pi\rangle$. Indeed the usual point-wise product for functions over Minkowski space are replace by a {\it non-commutative} $\star$-product 
\beq
e_{\pi_1}(x) \star e_{\pi_2}(x) \equiv e_{\pi_1\cdot \pi_2}(x)
\eeq
reflecting the non-abelian nature of the momentum group manifold $G$. The usual algebra of functions on $\mathbb{R}^{3,1}$ representing ``wave-functions" is now replaced by a non-commutative algebra and thus deformed quantum kinematics may be seen as ``non-commutative geometry" of the Hilbert space of quantum states. In analogy with the undeformed case the four-momentum observable $P^{\mu}$ associates four-vectors to the eigen-kets $|\pi\rangle$
\beq
P^{\mu} |\pi\rangle = \mathcal{P}^{\mu} (\pi) |\pi\rangle
\eeq
and thus $P^{\mu} e_{\pi}(x) =  \mathcal{P}^{\mu} (\pi)\, e_{\pi}(x)$. From this we see that 
\beq
P^{\mu}\, e_{\pi_1\cdot \pi_2}(x) = \mathcal{P}^{\mu} (\pi_1\cdot \pi_2) e_{\pi_1\cdot \pi_2}(x)
\eeq
and thus the total momentum of a quantum two-particle state is given by
\beq
\mathcal{P}^{\mu} (\pi_1\cdot \pi_2) = \mathcal{P}^{\mu} (\pi_1) \triangleright  \mathcal{P}^{\mu} (\pi_2)
\eeq
i.e. usual four-momenta addition $+$ is replaced by a non-abelian composition law $\triangleright$. Notoriously this non-abelian composition of momenta reflects in a non-Leibniz action of the quantum observable $P^{\mu}$. Indeed from
\beq
\langle \pi_1 \pi_2 | P^{\mu} |\pi_1 \pi_2 \rangle = \mathcal{P}^{\mu} (\pi_1) \triangleright  \mathcal{P}^{\mu} (\pi_2)
\eeq
one sees that 
\beq
P^{\mu} |\pi_1 \pi_2 \rangle \neq P^{\mu} |\pi_1\rangle \otimes |\pi_2\rangle +  |\pi_1\rangle \otimes P^{\mu} |\pi_2\rangle 
\eeq
but rather $P^{\mu}$ acts on tensor product states according to a deformed Leibniz rule which can be read off the non-abelian composition law and which we formally write
\beq
P^{\mu} (|\pi_1\rangle \otimes |\pi_2\rangle) = P^{(1)\mu} |\pi_1\rangle \otimes |\pi_2\rangle + |\pi_1\rangle \otimes P^{(2)\mu}  |\pi_2\rangle\, .
\eeq
This type of non-symmetric action on tensor product representations is typical of non-trivial Hopf algebras and in fact is the characterizing feature of deformations of the algebra of translation generators appearing both in the ``quantum double" and $\kappa$-Poincar\'e models. What we just showed is that such a non-trivial structure plays a role for multi-component quantum states but it {\it does not} affect the behaviour of classical observables.

\section{Summary}
In this work we formulated a general framework for consistently defining Poisson structures on phase spaces with group valued momenta. Borrowing tools from the theory of Poisson-Lie groups and Lie bi-algebras we showed that such structures can be constructed, using appropriate $r$-matrices, using as input just the algebraic structure of the generators of the momentum space Lie group. We applied our results to well studied examples of group momentum spaces reproducing and generalizing known results. Moving from single to multi-particle systems, we discussed the behaviour of four-momentum observables. We showed how single-particle momenta compose according to a ordinary sum to give the total four-momentum of a composite system.

The most pressing question, which is the subject of ongoing work, is how to generalize the picture we presented to include spin. Since the natural tool for including a relativistic particle's spin in the classical phase space is the use of the co-adjoint orbit method, we will look to adapt the existing extensions of the latter to Poisson-Lie groups to the specific framework of deformed phase-spaces. The other important question is how to bridge the picture we presented in this work with the associated quantum deformations of the Poincar\'e group and algebra. The key to this connection will be a judicious use of the Drinfeld double structure at the classical level. This will provide the classical $r$-matrices compatible with the momentum group structure which will yield the ``infinitesimal" structure to be connected with the deformed Leibniz rule associated to the deformation.  Upon quantization we expect all the structure of the deformed relativistic symmetries associated to the various momentum Lie group to emerge naturally from such classical structure.  This will allow to complete a picture in which deformed relativistic phase spaces and their associated quantum group symmetries emerge solely from the specification of the non-trivial Lie group structure of momentum space.


\section*{Acknowledgements}
We are grateful to A. Yu. Alekseev and F. Girelli for useful correspondence on the theory of Poisson-Lie groups and Lie bi-algebras.
The work of MA was supported by a Marie Curie Career Integration Grant within the 7th European Community Framework Programme and by the John Templeton Foundation. FN acknowledges support from CONACYT grant No. 250298.


\end{document}